\documentclass[preprint,11pt]{elsarticle}

\usepackage[left=3cm,right=3cm,bottom=3.5cm]{geometry}
\usepackage{subcaption}
\usepackage[font=small,labelfont=bf,tableposition=top]{caption}
\usepackage{graphicx}
\usepackage{bbold}
\usepackage{url}
\usepackage{booktabs}
\usepackage{color}
\usepackage{xcolor}
\usepackage[T1]{fontenc}

\usepackage{appendix}

\usepackage[hidelinks,breaklinks=true]{hyperref}
\usepackage{xurl}

\usepackage{hyperref}

\begin{document}
	
\makeatletter
\def\ps@pprintTitle{}
\makeatother

\begin{frontmatter}

\title{Ultrafast Extreme Events:\\Empirical Analysis of Mechanisms and Recovery \\in a Historical Perspective}

\author{Luca Henrichs}\ead{luca.henrichs@stud.uni-due.de}
\author{Anton J. Heckens}\ead{anton.heckens@uni-due.de}
\author{Thomas Guhr}\ead{thomas.guhr@uni-due.de}
\address{Fakult\"at f\"ur Physik, Universit\"at Duisburg--Essen, Duisburg, Germany}

\begin{abstract}
To understand the emergence of Ultrafast Extreme Events (UEEs), the influence of algorithmic trading or
high--frequency traders is of major interest as they make it extremely difficult to intervene and to stabilize financial markets.
In an empirical analysis, we compare various characteristics of UEEs over different years for the US stock market to assess the possible non--stationarity of the effects.
We show that liquidity plays a dominant role
in the emergence of UEEs and find a general pattern in their dynamics.
We also empirically investigate the after--effects in view of the recovery rate. We find common patterns for different years. We explain changes in the recovery rate by varying market sentiments for the different years.
Overall, our results hint at a certain degree of universal behavior.
\end{abstract}

\end{frontmatter}

\section{\label{sec:level1}Introduction}

Unusually strong and rapid price changes and the resulting instabilities may trigger major crises on the financial markets. Such events are referred to as Ultrafast Extreme Events (UEEs) or mini--flash crashes~\cite{sornette2011crashes_2011,golub2012highfrequencytradingmini,johnson2013abrupt,McInish_2014,biais2014hft,LalyPetitjean_2020,STEFFEN2023100806,desagre2024revisiting}.
In contrast to flash crashes, such as the May 6, 2010 Flash Crash~\cite{nanex2010a,cftc_sec2010a,cftc_sec2010b,foresight2012,Virgilio_2019}, UEEs last a few second~\cite{johnson2013abrupt}. They occur on exchanges worldwide \cite{johnson2013abrupt,felez2017market,aquilina2018participants,MESTEL2024111982}, and
their mechanisms are controversially discussed~\cite{Hofstetter,desagre2024revisiting}. 
In particular, high--frequency traders (HFTs), \textit{i.e.} special algorithmic traders~\cite{HENDERSHOTT_2011}, have been in the focus~\cite{golub2012highfrequencytradingmini,O_Hara2014,KIRILENKO_2017,braun2018impact,desagre2024revisiting}.
These HFTs are computer programs that autonomously trade assets on exchanges. 
In particular, several studies have focused on the role of HFTs in market liquidity~\cite{Easley_2012,CHORDIA_2013,Brogaard_2014}.
Due to their rapid data processing, they can react faster to UEEs than humans. The corresponding short time scales severely limit the possibility for human personnel to intervene and stabilize financial markets.
Hence, HFTs might cause UEEs or they might play a major role in their dynamics~\cite{ozenbas2018high,aquilina2018participants,desagre2019high,desagre2024revisiting}.
A different mechanism for the emergence of UEEs might be the behavior of human traders \cite{braun2018impact}. ``Large'' market orders might lead to large jumps in the bid--ask spread and thus trigger a UEE. A ``large`` market order is defined~\cite{braun2018impact} as a single market order which leads to one large return in the quotes (``large relative to quote price change'') during a UEE. In 2007/2008, 60$\%$ of UEEs were triggered by a ``large'' market order.
Other discussed mechanisms for the emergence of UEEs are market fragmentation~\cite{golub2012highfrequencytradingmini,felez2017market} and intermarket sweep orders (ISOs)~\cite{golub2012highfrequencytradingmini,johnson2013abrupt,shearer2020phases}.

As financial markets are in an extreme state during these events, it is also of interest to identify the time scales of the price recovery. After UEEs, full recovery after few trades as well as incomplete recovery after longer times has been observed~\cite{braun2018impact}.
Their rapid recovery is an indication that they most likely do not result due to the arrival of exogenous news~\cite{johnson2013abrupt}.
Furthermore, UEEs have a negative, decreasing impact on the liquidity of stocks \cite{golub2012highfrequencytradingmini}.

We present new results, extending the study in Ref.~\cite{braun2018impact} to the years 2014 and 2021, and we compare the findings over the course of time with the time period 2007/2008, \textit{i.e.} we study and compare the non--stationarity in the dynamics and recovery of UEEs from a period of human and machine trading to almost pure machine trading~\cite{kissell2020algorithmic}.
Here, ``non-stationarity”  refers to structural changes in market conditions across years, not to formal time-series stationarity.
We want to see if there is a general pattern in the emergence and dynamics of UEEs due to the liquidity as its relevant driving mechanism.
The consistency of the observed patterns shows a clear tendency towards universal behavior.
As indicated in the title, our approach is purely empirical; we explore statistical properties and do not rely on a model interpretation which would always involve assumptions beyond empirical validation.\\
In Sec.~\ref{sec:DataSet} we introduce our data set. We define UEEs in Sec.~\ref{sec:DefinitionUEEs} and show our results in Sec.~\ref{sec:Results}. We conclude in Sec.~\ref{sec:Conlcusion}.

\section{\label{sec:DataSet}Data Set}

Our study is based on the Daily TAQ (Trade and Quote) for the years 2014 and 2021~\cite{NYSE_20142021}.
The trade file contains information on trades and the corresponding time stamps and trading volumes.
The quote file lists the best bid and best ask prices.
We use the trade and quote file for the detection of UEEs and large returns in the quotes, respectively.
While the time resolution in 2014 is milliseconds, it is nanoseconds for 2021. In Ref.~\cite{braun2018impact}, the maximum time resolution was one second. In total, we study 100 high--liquid stocks for each year. In Ref.~\cite{braun2018impact}, all stocks of the S\&P 500 which were continuously traded during 2007 and 2008 were considered.
Due to liquidations, bankruptcy, mergers or acquisitions the selection of stocks for 2014 and 2021 is different. There are 81 stocks in common for both years.
The selected stocks for 2014 and 2021 are listed in App.~\ref{Appendix}.

\section{\label{sec:DefinitionUEEs}Definition of Ultrafast Extreme Events}

\begin{figure}[htbp]
	\centering 	
	\includegraphics[width=0.7\linewidth]{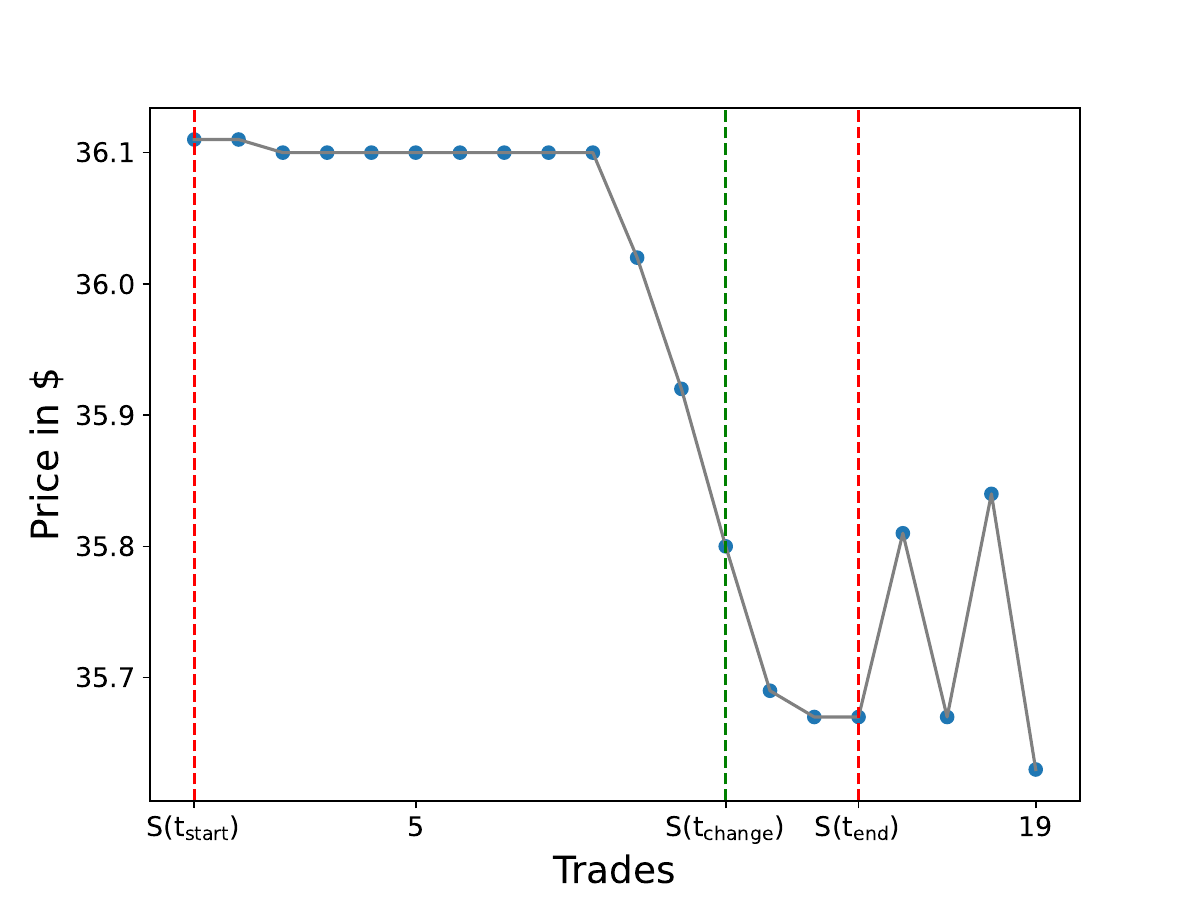}
	\caption{Beginning and end of a UEE with prices $S{(t_\mathrm{start})}$ and $S{(t_\mathrm{end})}$ for \textsc{Microsoft} on 7 January 2014. The price at which the $0.8\%$ criterion is first fulfilled is denoted by $S{(t_\mathrm{change})}$.}
	\label{Beispiel}
\end{figure}
According to \textsc{Nanex}~\cite{Nanex}, a UEE is defined as a monotonic price change where the price changes by more than 0.8\% over more than ten trades and less than 1.5 seconds.
In the sequel, we employ the definition by \textsc{Nanex} to identify UEEs.
We note that there is a definition for related events referred to as extreme price movements where the 99.9th percentile of the absolute values for ten second returns are used \cite{BROGAARD2018253}.
An upward price trend is referred to as flash spike, a downward one as flash crash. We emphasize that we exclude events lasting longer than 1.5 seconds.
The UEE ends when the increasing or decreasing monotonic trend of price changes for the flash spike or the flash crash ends, respectively.
Figure~\ref{Beispiel} shows a flash crash. We denote the 
price at the start of a UEE by $S{(t_\mathrm{start})}$ and the price at the end of a UEE by $S{(t_\mathrm{end})}$. The price at which the $0.8\%$ criterion is fulfilled for the first time is denoted by $S{(t_\mathrm{change})}$.

In addition to the 1.5-second criterion, we also use two seconds as the duration criterion for a UEE.
There are two reasons for this. 
First, we want to test the robustness of the criterion.
Second, we will refer to results from Ref.~\cite{braun2018impact}, where two seconds were used instead of the 1.5 seconds.

\section{\label{sec:Results}Results}

In Sec.~\ref{2021}, we present statistical aspects of UEEs for 2014 and 2021.
In Sec.~\ref{2014}, we study the role of liquidity in the formation of UEEs. The after--effects of UEEs are analyzed in Sec.~\ref{sec:Impact} via the recovery ratio.

\subsection{Statistical Aspects of Ultrafast Extreme Events for 2014 and 2021} \label{2021}

\begin{figure}[htbp]
	\centering
	\centering
	\includegraphics[width=1\textwidth]{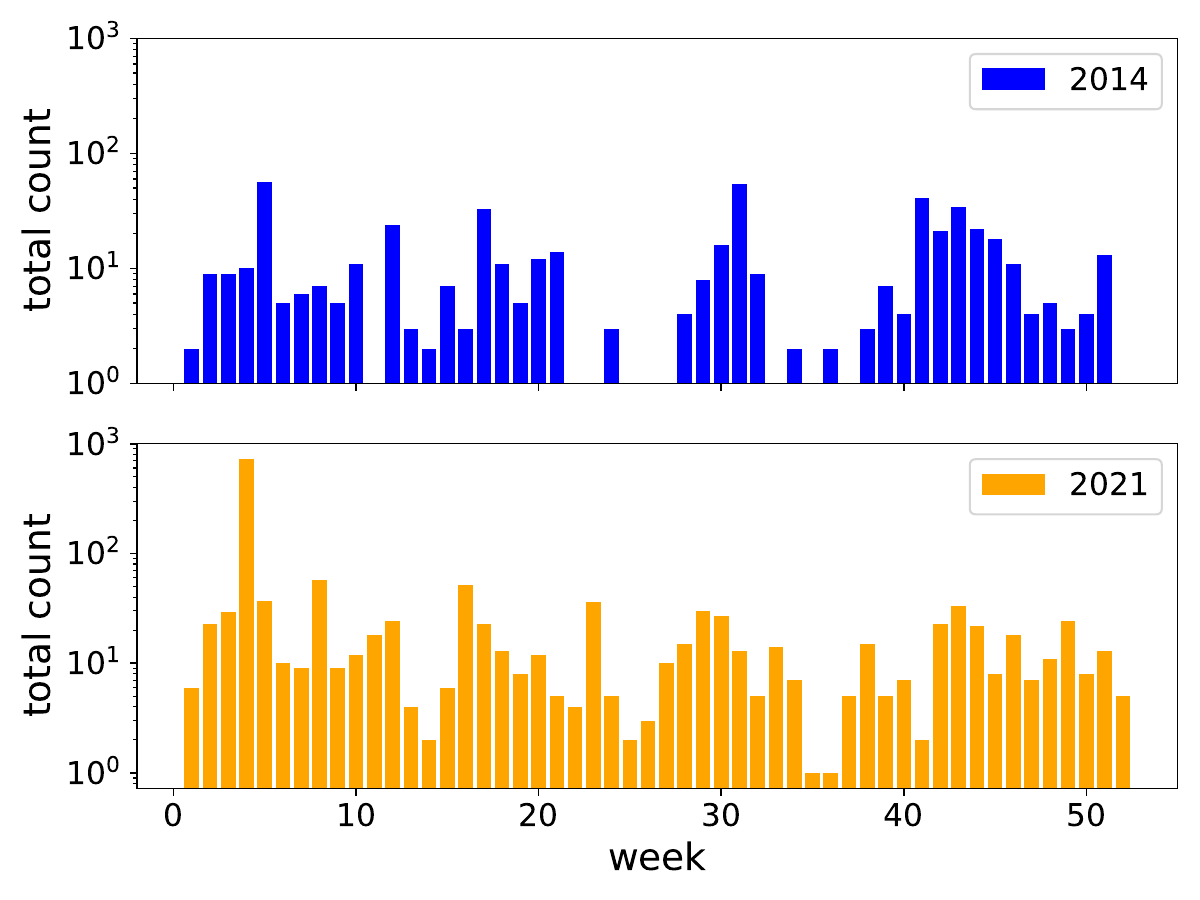}
	\caption{Total number of counts for UEEs on a logarithmic scale for every trading week in 2014 (top), in 2021 (bottom).}
	\label{Jahreshis. 2014}
\end{figure}

Figure~\ref{Jahreshis. 2014} shows the total number of counts for UEEs in 2014 and 2021. Additionally, we list the number of UEEs in Tab.~\ref{NumberUEEs} for both the 1.5- and the 2--second criteria.
\begin{table}[htbp]
	\centering
	\caption{Overview of number of UEEs, spikes, and crashes in 2014 and 2021 for different UEE criteria.}
	\label{NumberUEEs}
	\begin{tabular}{@{}ccccc@{}}
		\toprule
		Year & \parbox[t]{2.5cm}{\centering UEE\\criterion} & \parbox[t]{2.5cm}{\centering Number\\of UEEs} & Spikes & Crashes \\
		\midrule
		2014 & 1.5s  & 526  & 244 (46.4\%) & 282 (53.6\%) \\
		2014 & 2.0s  & 585  & 273 (47.0\%) & 312 (53.0\%) \\
		2021 & 1.5s  & 1461 & 650 (44.5\%) & 811 (55.1\%) \\
		2021 & 2.0s  & 1564 & 708 (45.0\%) & 856 (55.0\%) \\
		\bottomrule
	\end{tabular}
\end{table}
The 2--second criterion yields few additional UEEs of the order less than 12\,\%, suggesting that most are completed within 1.5 seconds.
We find that UEEs rarely occur individually but are often clustered, \textit{i.e.} when several UEEs are counted in one day, the following days often show a similar picture.
For 2021, we notice clustering of UEEs as well.
The clustering of UEEs can be interpreted in terms of short-term order-book instability. In market microstructure it is well established that temporary reductions in order-book depth and liquidity resilience tend to persist over short intervals, thereby increasing the likelihood of additional abrupt price movements within the same period \cite{Bouchaud2009_3,Hasbrouck2007,FarmerLillo2004}. Such short-lived liquidity shortages naturally generate clusters of extreme events. This microstructural interpretation is also consistent with broader frameworks of volatility clustering, including those discussed in the Fractal Market Hypothesis \cite{Peters1994}.

In 2014, the maximum number of UEEs found in one day is 33.
In contrast to 2014, there is a significant increase in UEEs in January, more precisely in the period 25 January to 28 January 2021. The largest number of UEEs is counted on 28 January with 492 events, 444 of these occurred in the stock of \textsc{American Airlines}. This increase in UEEs can be explained by a Short Squeeze Event in January 2021~\cite{enwiki:1269767108}.

\begin{figure}[htbp]
	\centering 	
	\includegraphics[width=1.0\linewidth]{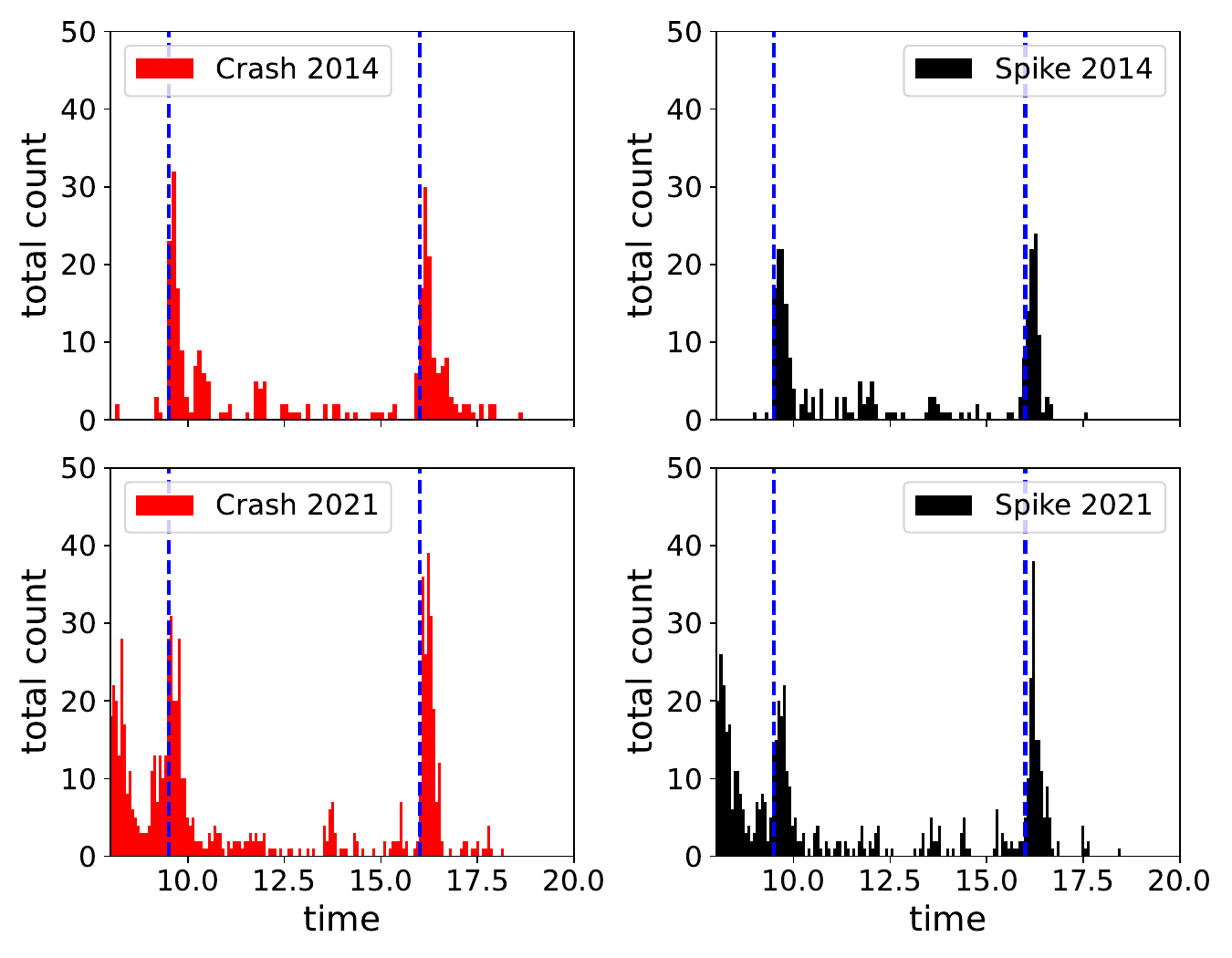}
	\caption{Frequency histograms for UEEs over a whole trading day for 2014 (top) and 2021 (bottom). The dashed lines indicate the start of the main phase of trading at 9:30 a.m. and its end at 4:00 p.m.}
	\label{fig:FrequencyUEEsForTradingDays}
\end{figure}

We study UEEs for the whole trading day including pre-- and after--market trading, see Fig.~\ref{fig:FrequencyUEEsForTradingDays}. Most UEEs occur after the start of the main phase of trading (9:30 a.m) and after its end (4:00 p.m.).
These patterns  hint at commonalities in how UEEs occur and recover.	

\subsection{Mechanism for Ultrafast Extreme Events} \label{2014}

\begin{figure}[htbp]
	\centering
	\centering
	\begin{subfigure}[b]{0.48\textwidth}
	\centering
	\includegraphics[width=1\textwidth]{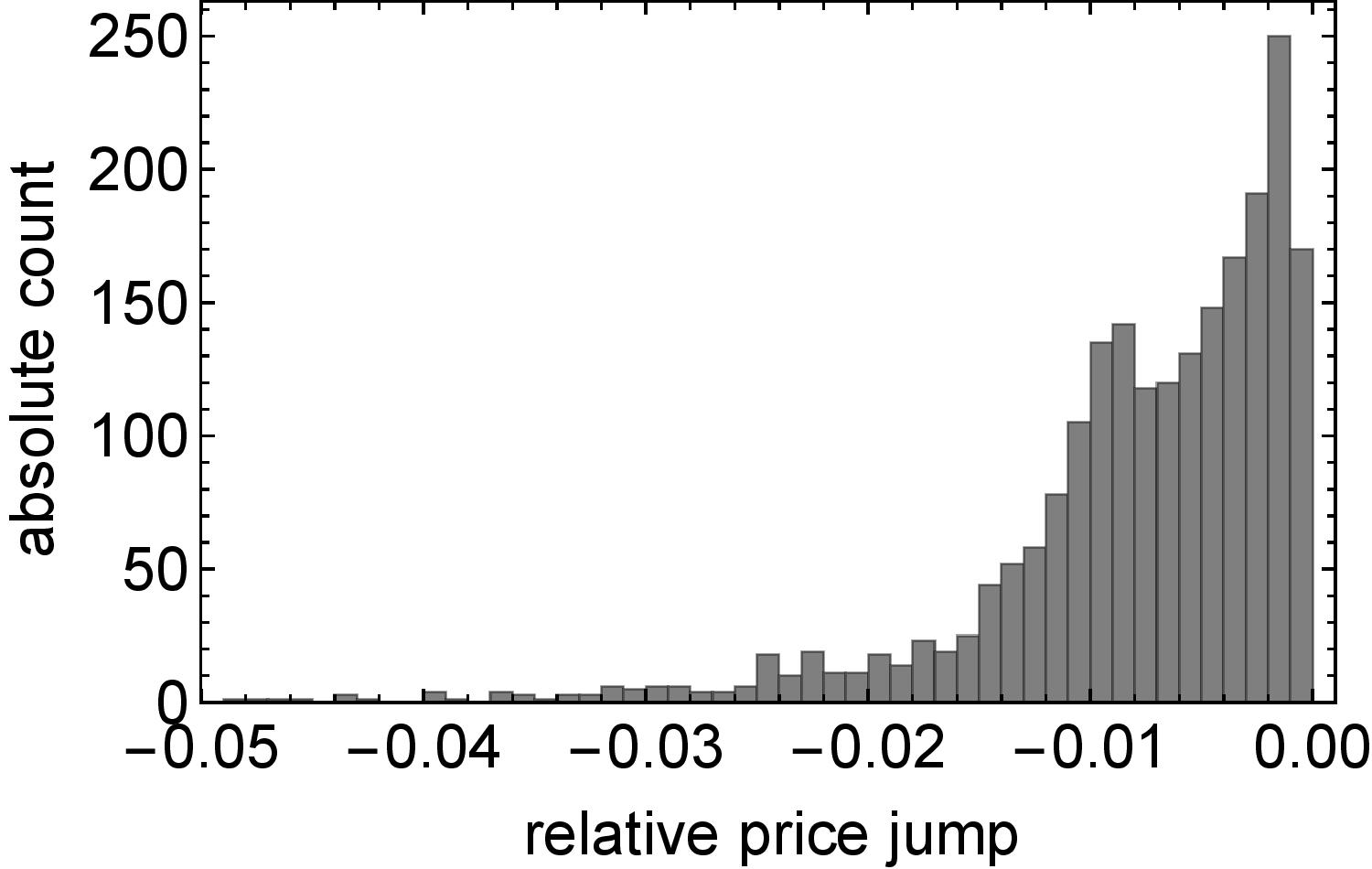}

\end{subfigure}
\hfill
\begin{subfigure}[b]{0.48\textwidth}
	\centering
	\includegraphics[width=1\textwidth]{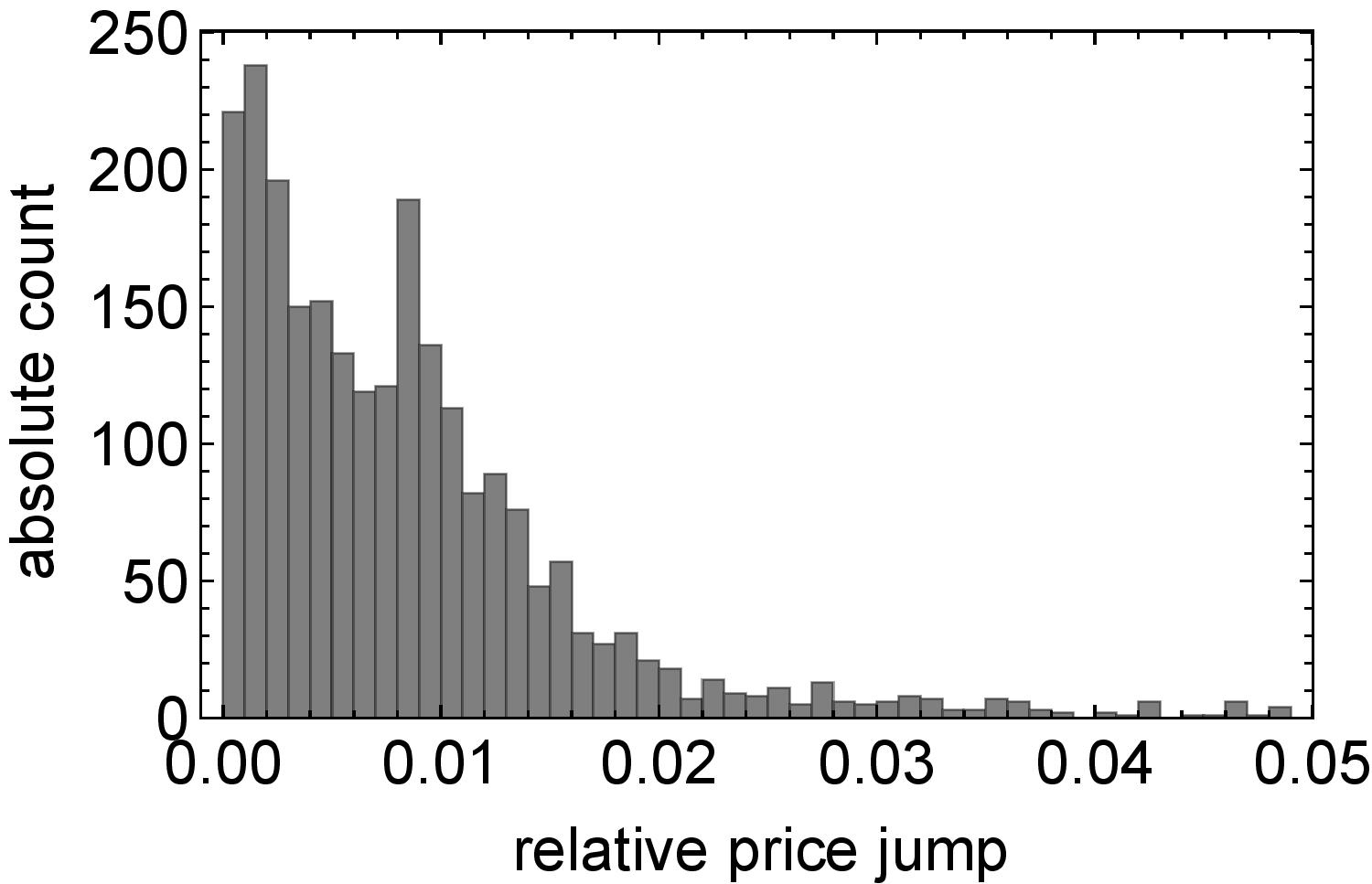}

\end{subfigure}\\
	\begin{subfigure}[b]{1.0\textwidth}
	\centering
	\includegraphics[width=1\textwidth]{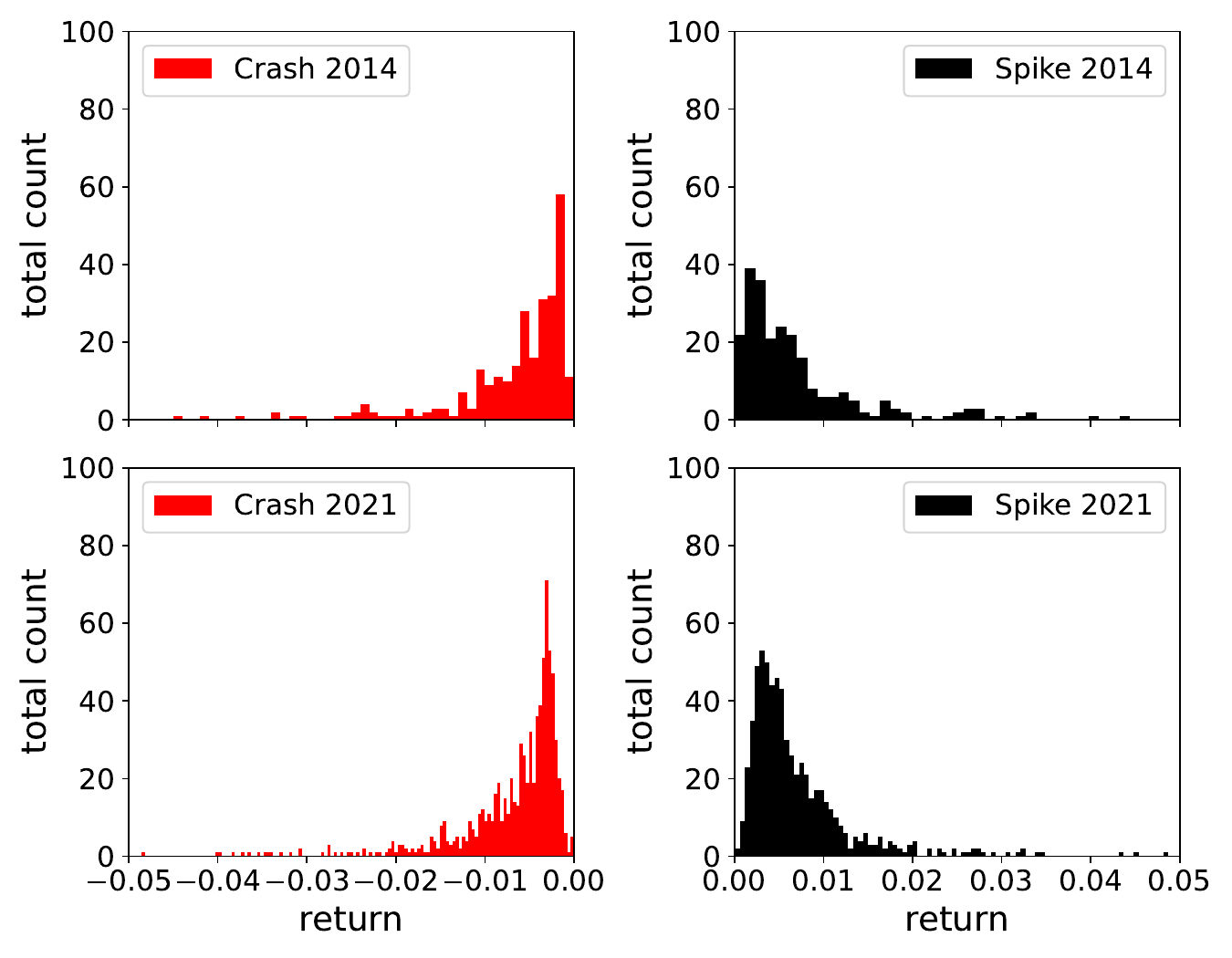}

	\end{subfigure}
	\caption{Total number of counts for the largest return in the quotes for flash crashes (left) and flash spikes (right) on a linear scale for 2007/2008 (top), 2014 (middle) and 2021 (bottom). The top figure is taken from Ref.~\cite{braun2018impact}.}
	\label{Bid jump 2014}
\end{figure}

\begin{figure}[htbp]
	\centering
	\begin{subfigure}[b]{0.7\textwidth}
		\centering
		\includegraphics[width=1\textwidth]{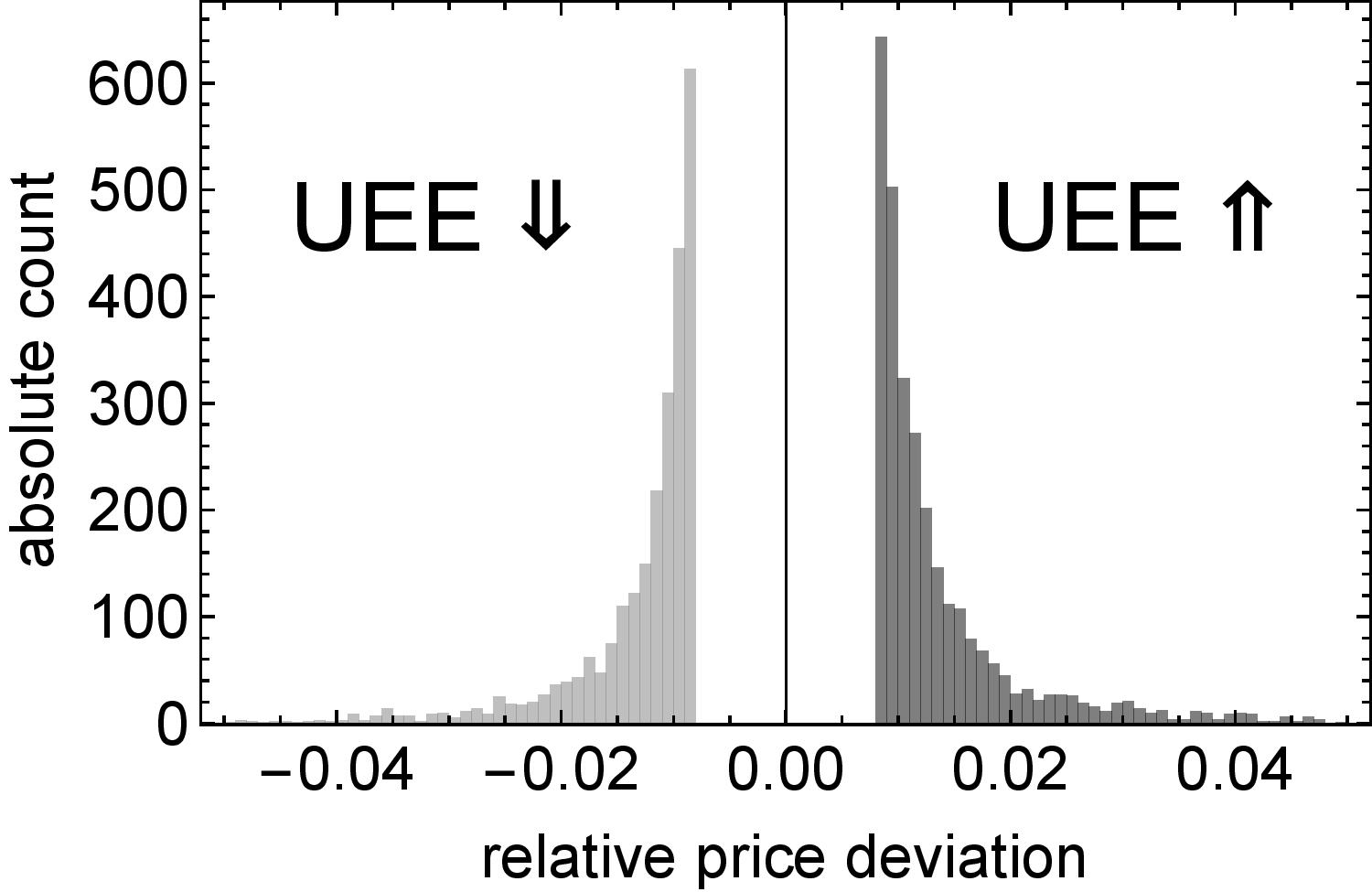}
	\end{subfigure}
	\begin{subfigure}[b]{1.0\textwidth}
		\centering
		\includegraphics[width=1\textwidth]{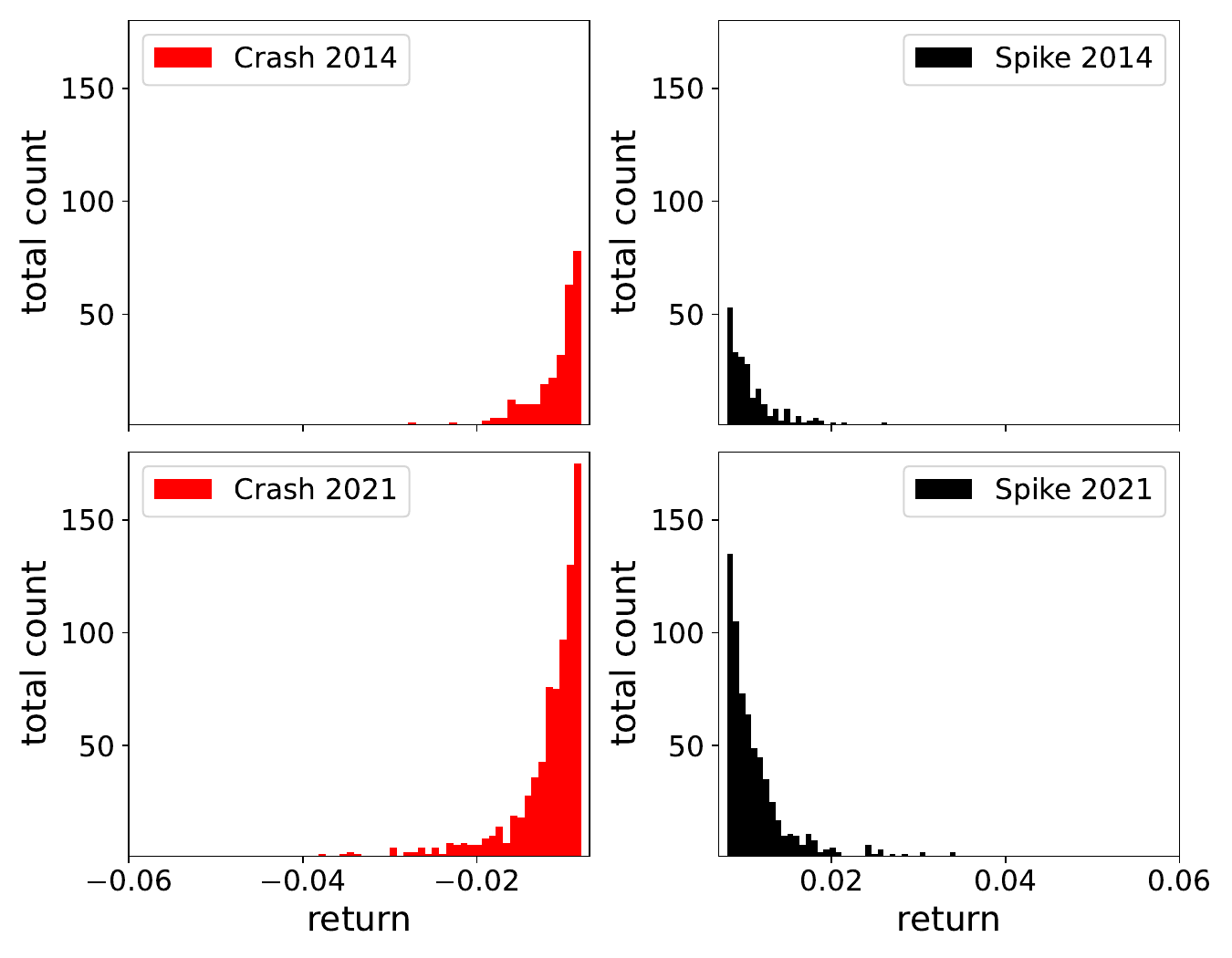}
	\end{subfigure}
	\caption{Frequency histograms with the total number of counts of the return values for flash crashes (left) and flash spikes (right) on a linear scale for 2007/2008 (top), 2014 (middle) and 2021 (bottom). The top figure is taken from Ref.~\cite{braun2018impact}.}
	\label{Preis 2014}
\end{figure}

We determine the largest return in the best bid for flash crashes and best ask for flash spikes within the period from the start to the end of a UEE~\cite{braun2018impact}. 
Figure~\ref{Bid jump 2014} shows the frequencies of the largest return in the quotes referred to as ``relative price jumps'' in Ref.~\cite{braun2018impact} during a UEE for the combined years 2007/2008, 2014 and 2021.
\begin{figure}[htbp]
	\centering
	\begin{subfigure}[b]{0.7\textwidth}
		\centering
		\includegraphics[width=1.0\textwidth]{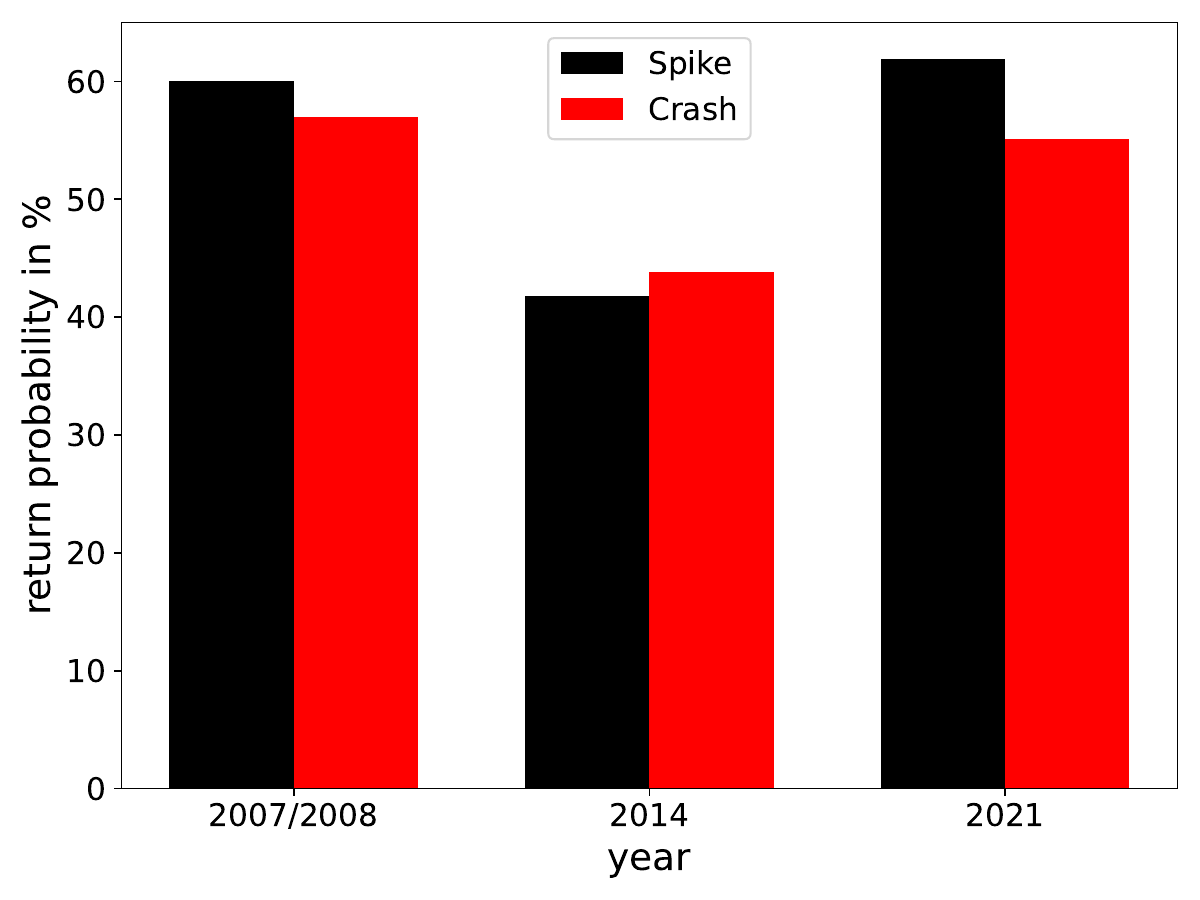}

	\end{subfigure}\\
	\begin{subfigure}[b]{0.7\textwidth}
		\centering
		\includegraphics[width=1.\textwidth]{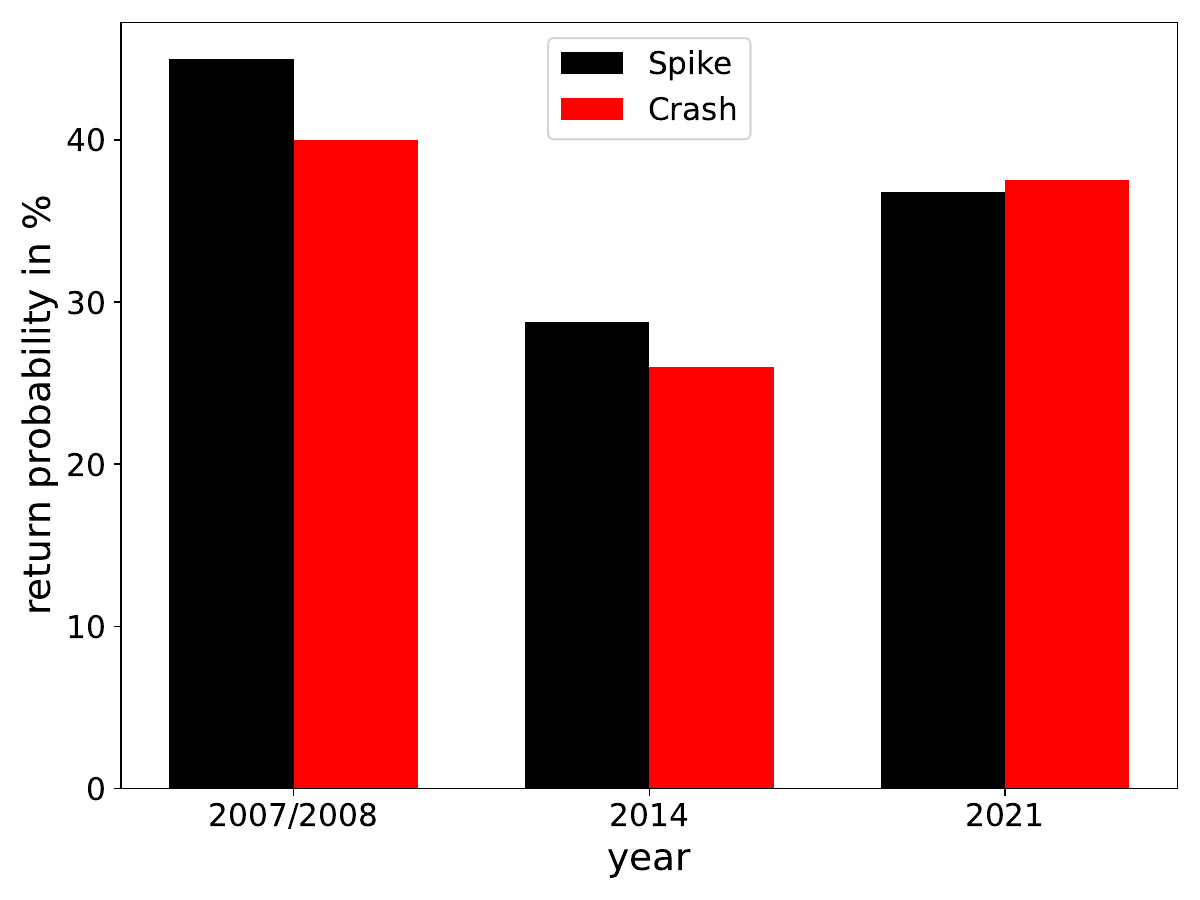}

	\end{subfigure}
	\caption{Barplots for absolute returns in the quotes larger than 0.005 (top) and 0.008 (bottom).
		Numbers for 2007/2008 are taken from Ref.~\cite{braun2018impact}.}
	\label{Barplot_ReturnInQuotes}
\end{figure}
We are interested in events that approximately fulfill the 0.8\,\% criterion by \textsc{Nanex}.
Hence, we show the absolute returns in the quotes larger than 0.005 and 0.008 in Fig.~\ref{Barplot_ReturnInQuotes}.
Between 40\% to 62\% of the absolute returns in the quotes are larger than 0.005 across all years.
Furthermore, we focus on the returns of a UEE
referred to as ``relative price deviations'' in Ref.~\cite{braun2018impact} defined as
\begin{eqnarray}
	r_{\mathrm{UEE}} = \frac{S{(t_\mathrm{end})} - S{(t_\mathrm{start})}}{S{(t_\mathrm{start})}} \,.
\end{eqnarray}
Figure~\ref{Preis 2014} shows the frequency histograms of the returns for UEEs for 2014 and 2021. 
Comparing both the distributions of returns for UEEs and the distributions of the largest returns in the quotes, we find that many of the largest returns in the quotes are in magnitude comparable to the UEE returns, \textit{i.e.} they are also often larger than the 0.8\% criterion by \textsc{Nanex}. Thus, the large returns in the quotes are an important feature for many UEEs over the years.

\begin{figure}[htbp]
	\centering
	\begin{subfigure}[b]{1.0\textwidth}
		\centering
		\includegraphics[width=1.0\textwidth]{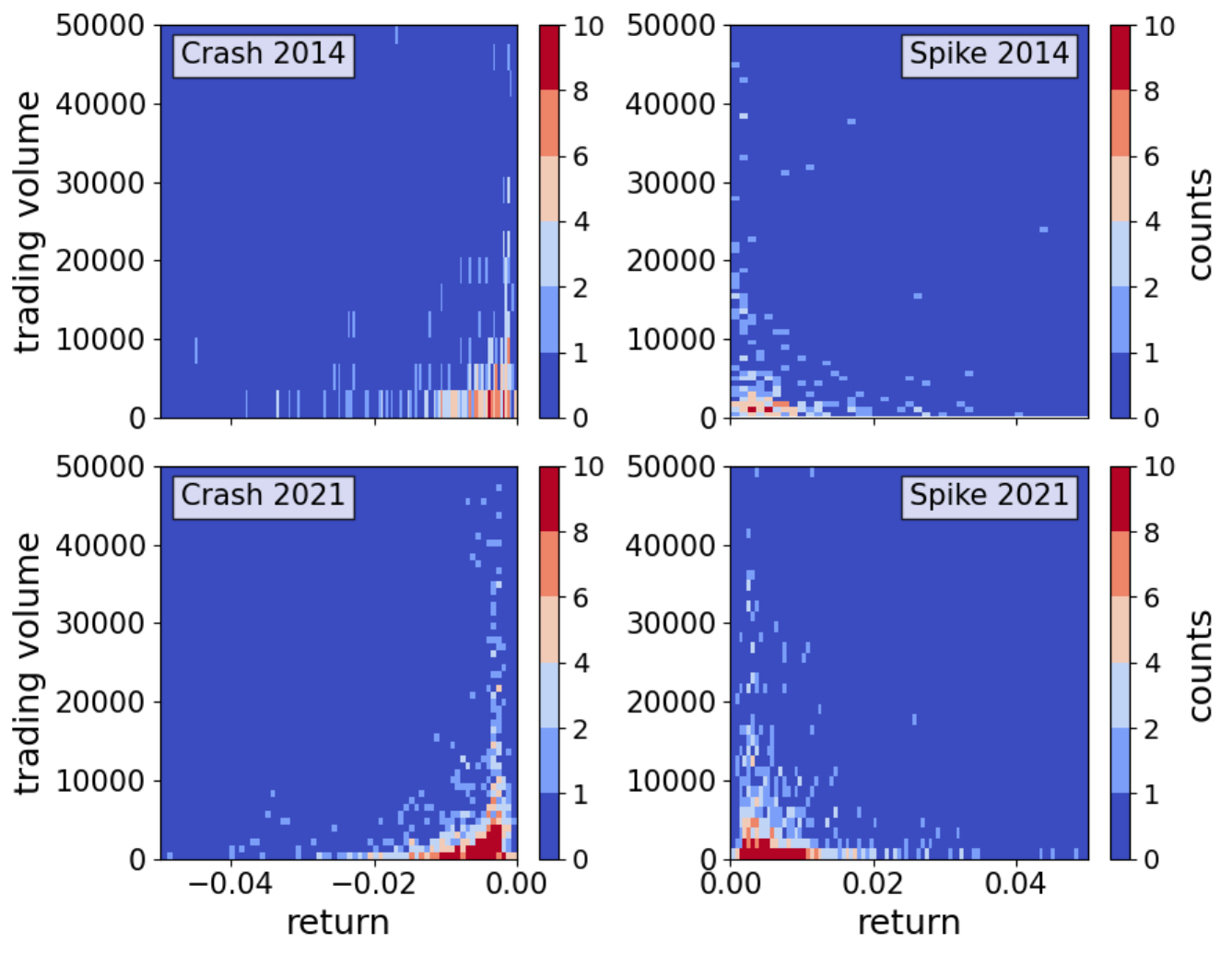}
	\end{subfigure}
	\caption{Two--dimensional frequency histograms for the accumulated trading volume and the largest return in the quotes for crashes (left) and spikes (right) in 2014 (top) and 2021 (bottom).}
	\label{2d 2014}
\end{figure}

Next, we determine the accumulated trading volume corresponding to the largest return in the quotes
by summing up all trading volumes 
from the start of a UEE to the trade at which the UEE criterion of 0.8\% price change is fulfilled, see Sec.~\ref{sec:DefinitionUEEs}.
Figure~\ref{2d 2014} shows for each largest return in the quotes
the corresponding %
accumulated trading volume in a two--dimensional frequency histogram. 
We notice that larger accumulated trading volumes tend to correspond to smaller returns,
while the smaller accumulated trading volumes correspond to larger returns.
This prompts our conclusion that the most relevant driving mechanism is the liquidity of the stocks.
	
\begin{figure}[htbp]
	\centering
	\includegraphics[width=1.0\textwidth]{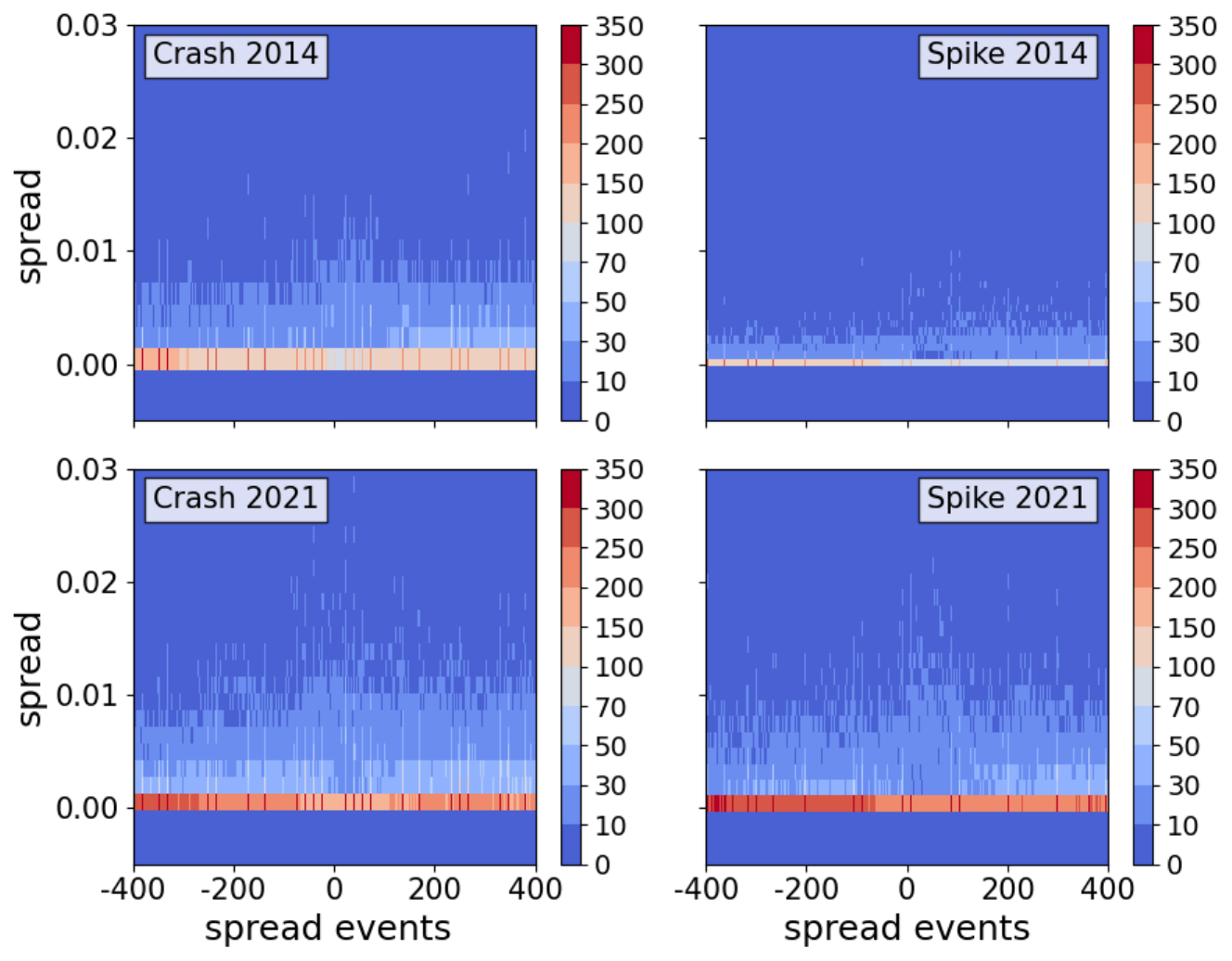}

	\caption{Contour plots of the relative spread, 400 spread events before and after a UEE begins for flash crashes (left) and flash spikes (right) in 2014 (top) and 2021 (bottom). Zero indicates the start of a UEE according to \textsc{Nanex}, see~Sec.~\ref{sec:DataSet}.}
	\label{ContourPlotSpread}
\end{figure}

To check this, we consider the spread before and after a UEE occurs, cf.~Ref.~\cite{golub2012highfrequencytradingmini}. To compare spreads of different stocks with each other, we calculate a relative spread
\begin{equation}
	s(t) = \frac{a(t) - b(t)}{a(t)}
\end{equation}
with best ask $a(t)$ and best bid $b(t)$.
A large relative spread occurs for low liquidity and a smaller one for a higher level of liquidity.
In Fig.~\ref{ContourPlotSpread}, we show the relative spreads for all stocks before and after a UEE.
Spread event zero indicates the start of a UEE according to \textsc{Nanex}, see~Sec.~\ref{sec:DefinitionUEEs}.
\begin{figure}[htbp]
		\centering
		\includegraphics[width=1.0\textwidth]{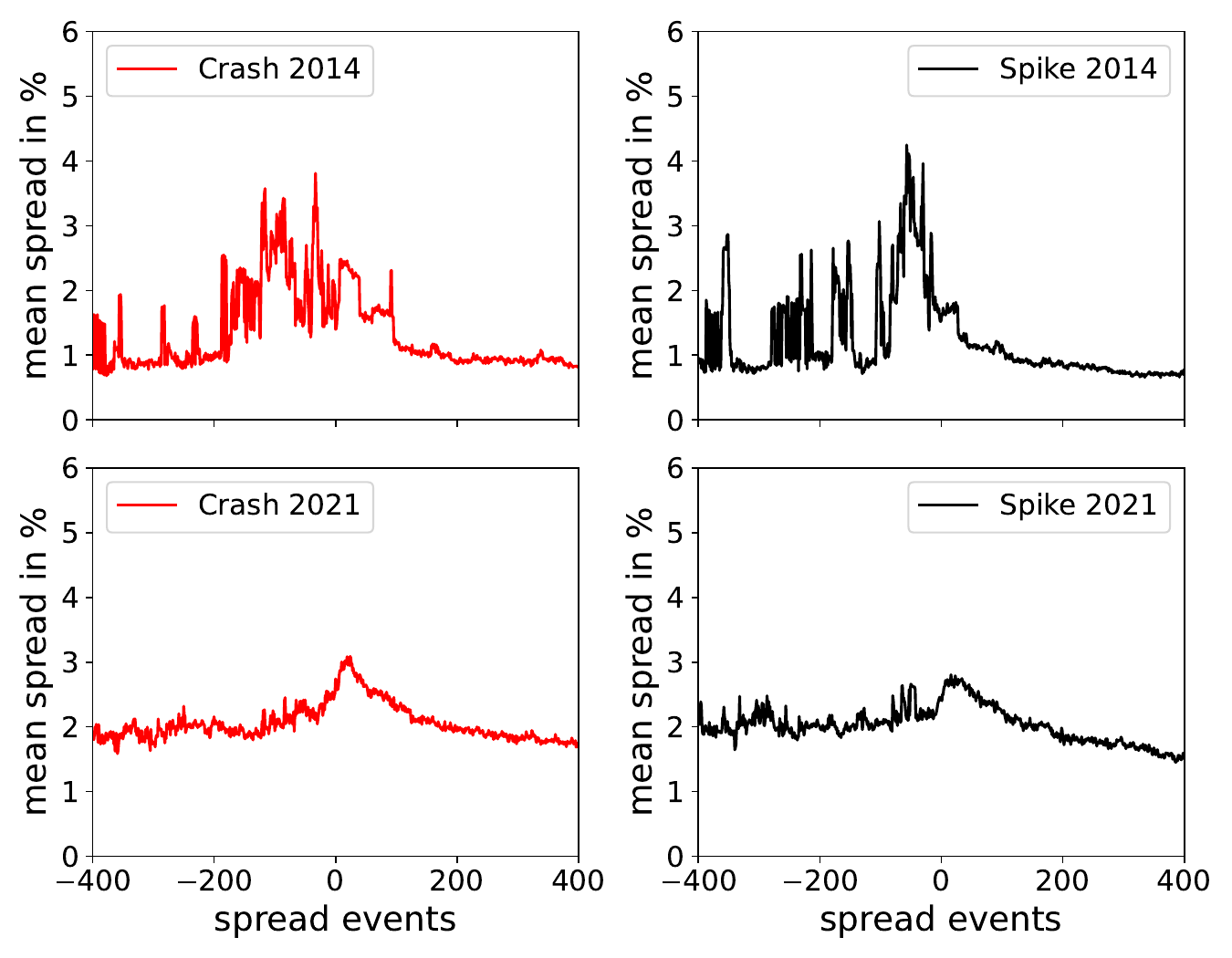}
	\caption{Relative spread, 400 spread events before and after a UEE begins, averaged across all UEEs for flash crashes (left) and flash spikes (right) in 2014 (top) and 2021 (bottom). Spread event zero indicates the start of a UEE according to \textsc{Nanex}, see~Sec.~\ref{sec:DataSet}.}
	\label{Spread 2014}
\end{figure}%
\begin{figure}[htbp]
	\centering
	\includegraphics[width=1.0\textwidth]{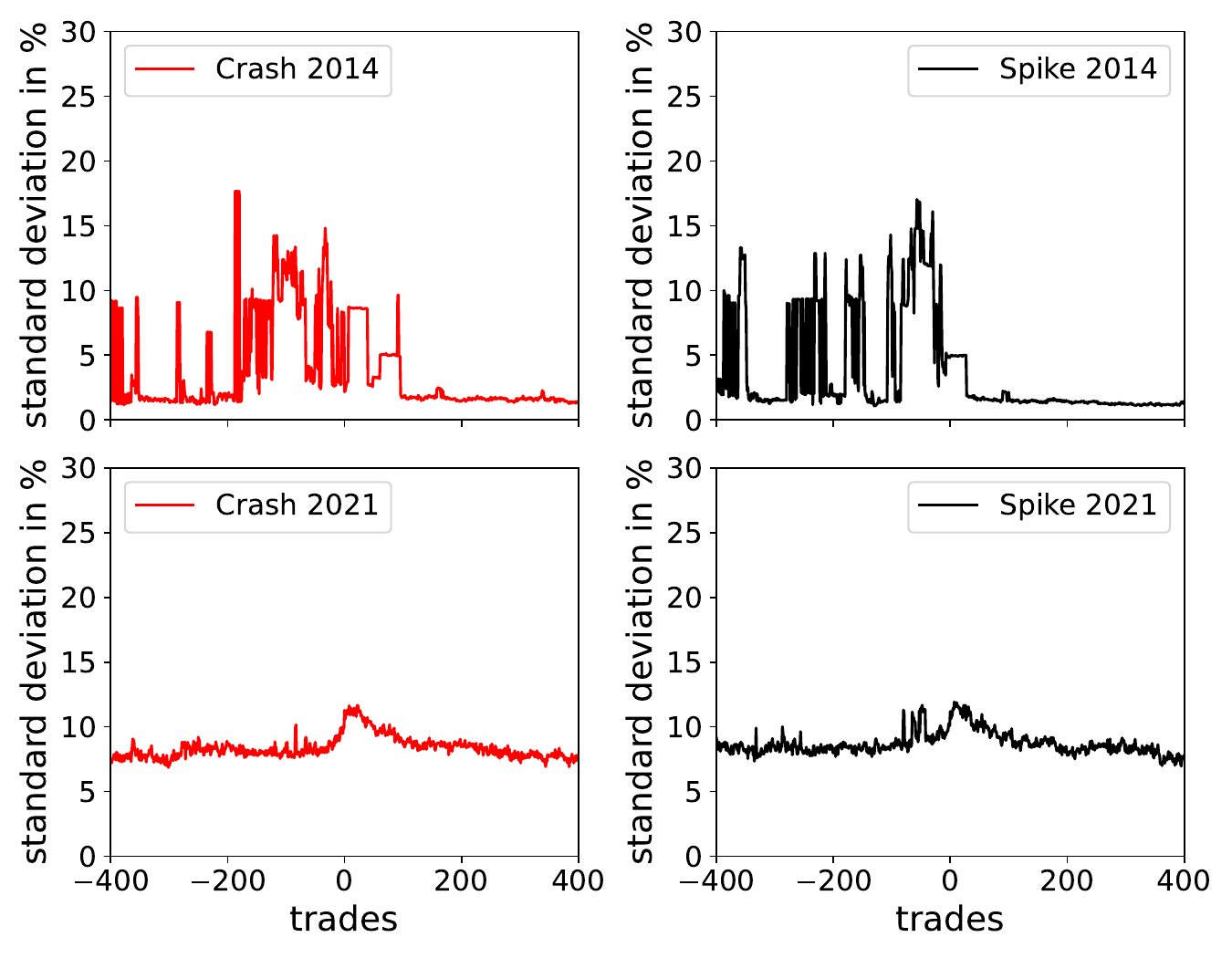}
	\caption{Standard deviations of relative spread, 400 spread events before and after a UEE begins, computed across all UEEs for flash crashes (left) and flash spikes (right) in 2014 (top) and 2021 (bottom). Spread event zero indicates the start of a UEE according to \textsc{Nanex}, see~Sec.~\ref{sec:DataSet}.}
	\label{Spread 2014_SD}
\end{figure}%
\begin{figure}[htbp]
	\centering
	\includegraphics[width=1.0\textwidth]{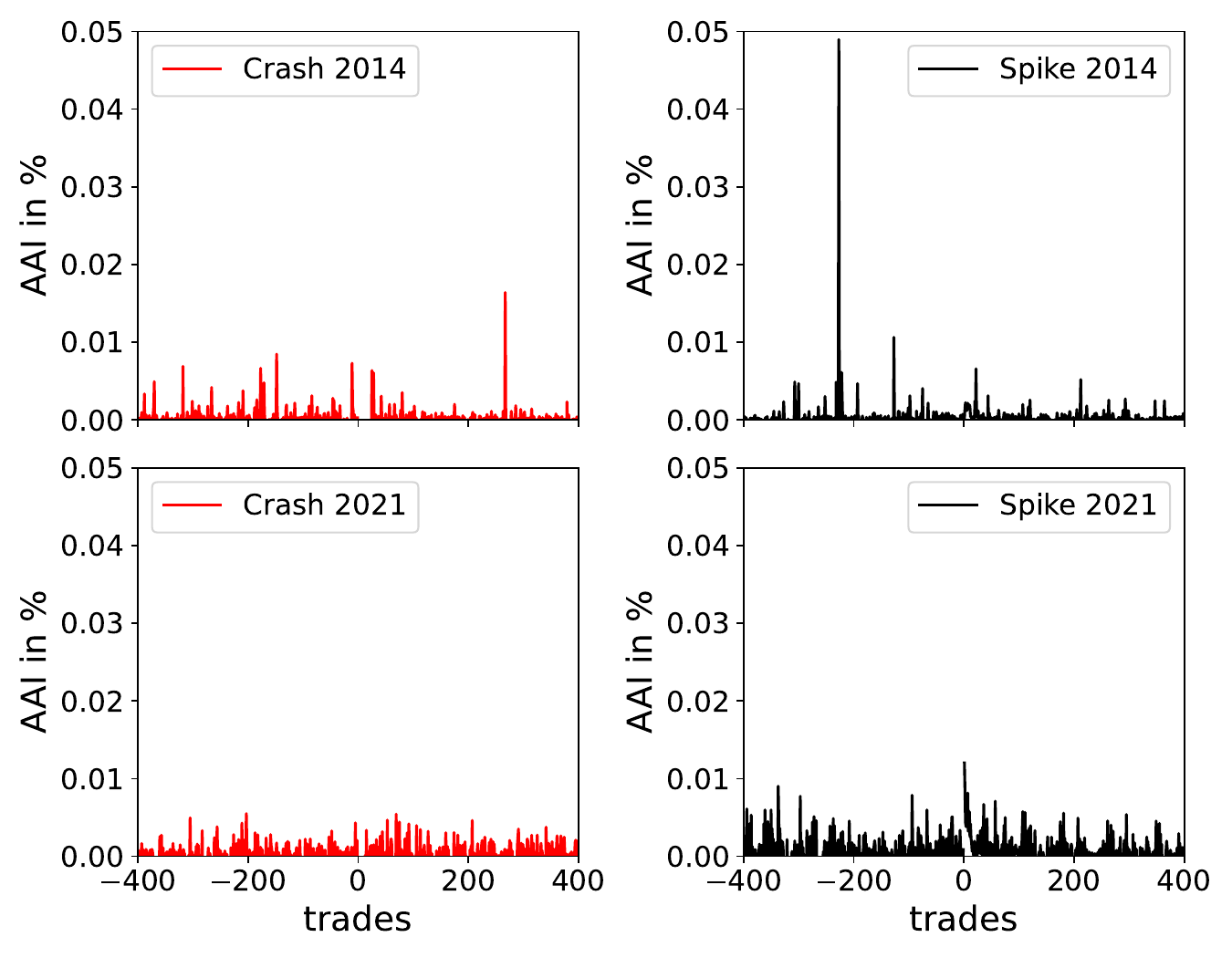}
	\caption{Adjusted Amihud illiquidity (AAI) of  400 spread events before and after a UEE begins, averaged across all UEEs for flash crashes (left) and flash spikes (right) in 2014 (top) and 2021 (bottom). Spread event zero indicates the start of a UEE according to \textsc{Nanex}, see~Sec.~\ref{sec:DataSet}.}
	\label{Amihud}
\end{figure}%
The relative spread
increases before the start of a UEE. 
Additionally, we notice large deviations between different stocks.
To make a trend more visible, we average over the different spreads for a fixed spread event in Fig.~\ref{Spread 2014}.
The corresponding standard deviations of the relative spread, computed across all UEEs for flash crashes and flash spikes, are depicted in Fig.~\ref{Spread 2014_SD}.
The averaged relative spread reveals a clear systematic trend.
Here, the increase in liquidity before the start of a UEE is more clearly visible.
This is consistent with the aforementioned observation that the largest returns in the quotes are associated with 
relatively small accumulated trading volumes.
We emphasize that UEEs can occur even in usually liquid stocks
when the liquidity drops momentarily.
Furthermore, we study
Amihud’s illiquidity ratio as one of the most widely used liquidity measures, as it combines price impact and trading activity into a single daily indicator~\cite{AMIHUD200231}. 
Here, we introduce an adjusted form of Amihud’s illiquidity (AAI)
\begin{equation}
	l_{AAI}(t) = \frac{|r(t)|}{v_{\mathrm{avg}}(t)} \, ,
\end{equation}
where $|r(t)|$ denotes the absolute value of consecutive trade-to-trade returns and $v_{\mathrm{avg}}(t)$ is the average traded volume computed over the corresponding consecutive trades.
		However, applying the classical Amihud logic at millisecond resolution is not meaningful: at such ultrafast timescales, trade-to-trade returns are dominated by discreteness, bid–ask bounce, and mechanical order-book effects.
		Figure~\ref{Amihud} clearly illustrates that the adjusted Amihud illiquidity measure, averaged across all UEEs, exhibits a fundamentally different time-scale behavior than the averaged relative spread in Fig.~\ref{Spread 2014}, further confirming that it cannot serve as a microstructural liquidity proxy at UEE timescales.

\subsection{\label{sec:Impact}Recovery of Stocks after Ultrafast Extreme Events}
To better understand the recovery process after UEEs on stock markets, we study the recovery ratio (referred to as recovery rate in Ref.~\cite{braun2018impact}),
defined by
\begin{equation}
	\eta_n = \frac{S(t_{\mathrm{end}}) - S(t_n)}{S(t_{\mathrm{end}}) - S(t_{\mathrm{start}})} \, ,
\end{equation}
where $S(t_{\mathrm{end}})$ is the price at the end of a UEE, $S(t_n)$ is the price at the $n$--th trade after the end of the UEE and $S(t_{\mathrm{start}})$ is the price at the beginning of the UEE. Hence, a recovery ratio of $\eta_n$ = 0 for $S(t_n) = S(t_{\mathrm{end}})$ indicates that the price is not recovered while
$\eta_n$ = 1 for $S(t_n) = S(t_{\mathrm{start}})$ shows that the stock price fully recovered. 

\begin{figure}[htbp]
	\centering
	\begin{subfigure}[b]{1.0\textwidth}
		\centering
		\includegraphics[width=1.0\textwidth]{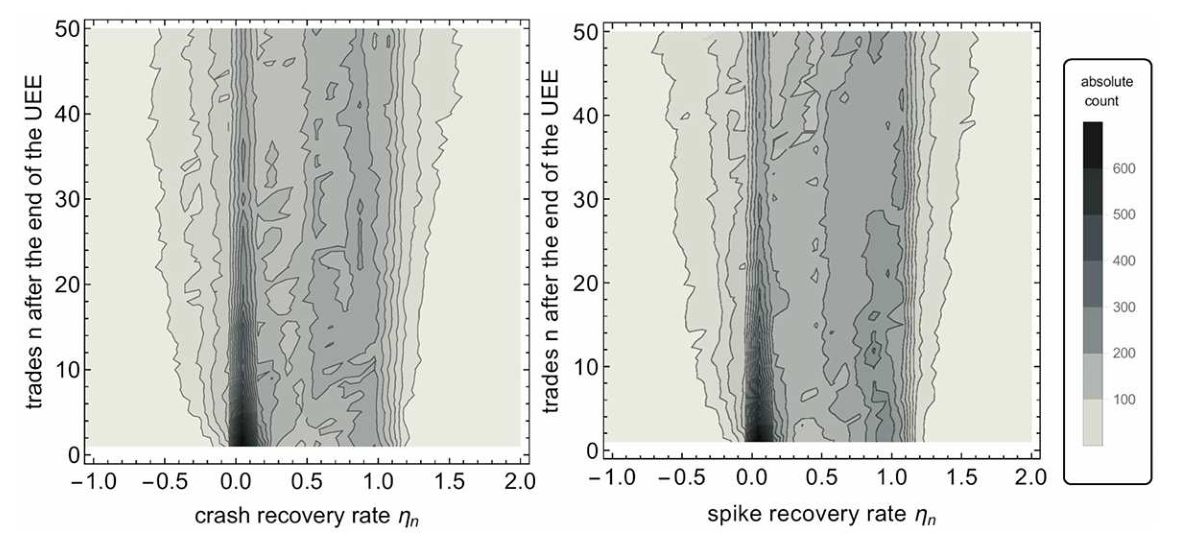}

	\end{subfigure}\\
	\begin{subfigure}[b]{1.0\textwidth}
		\centering
		\includegraphics[width=1.\textwidth]{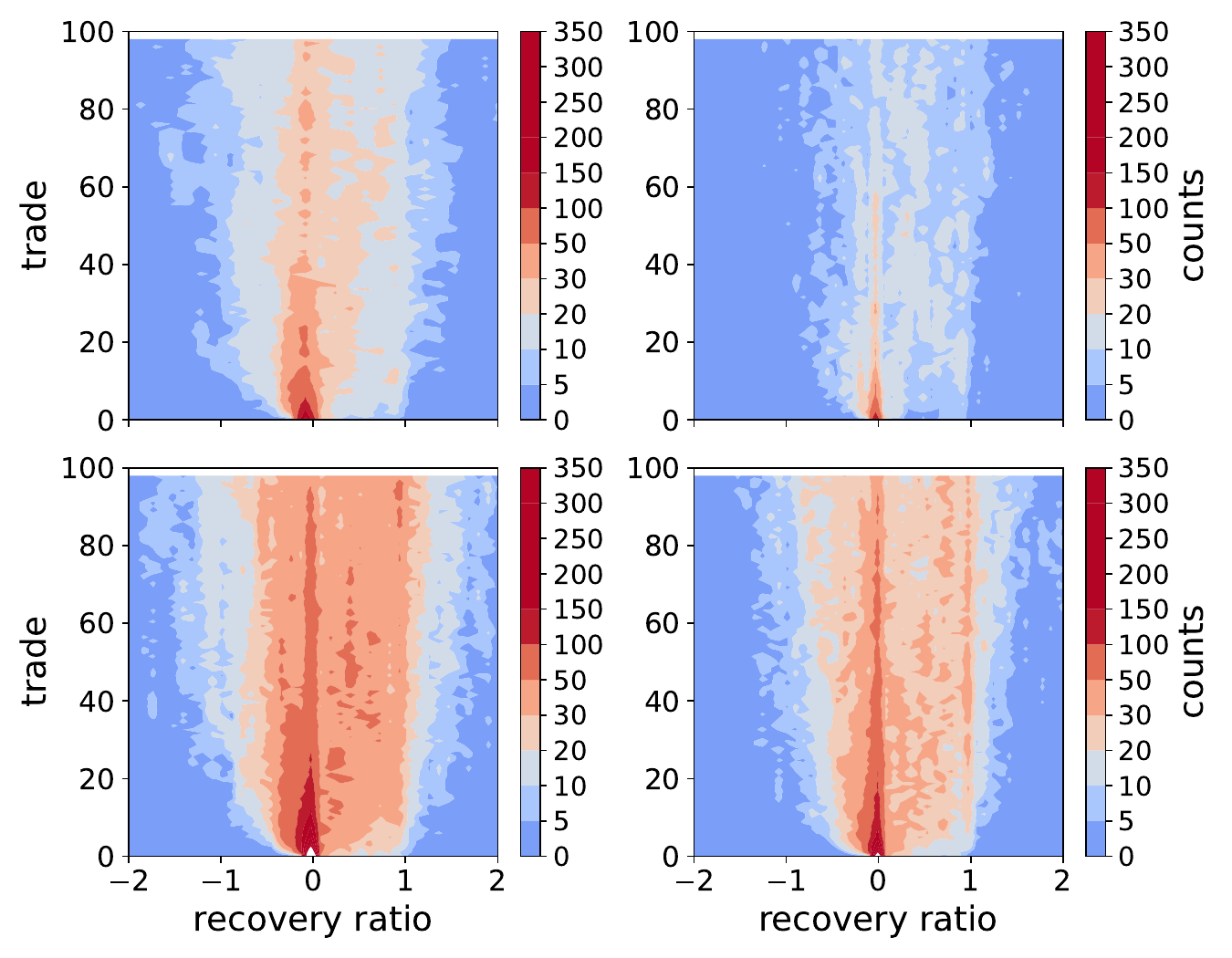}

	\end{subfigure}
	\caption{Contour plots of the recovery ratio (referred to as recovery rate in Ref.~\cite{braun2018impact}) dependent on the trade number for flash crashes (left) and flash spikes (right)
	in 2007/2008 (top), in 2014 (middle) and 2021 (bottom) on a linear scale.
	The top figure is taken from Ref.~\cite{braun2018impact}}.
	\label{Recover 2014}
\end{figure}

\begin{figure}[htbp]
	\centering
	\begin{subfigure}[b]{1.0\textwidth}
		\centering
		\includegraphics[width=1.\textwidth]{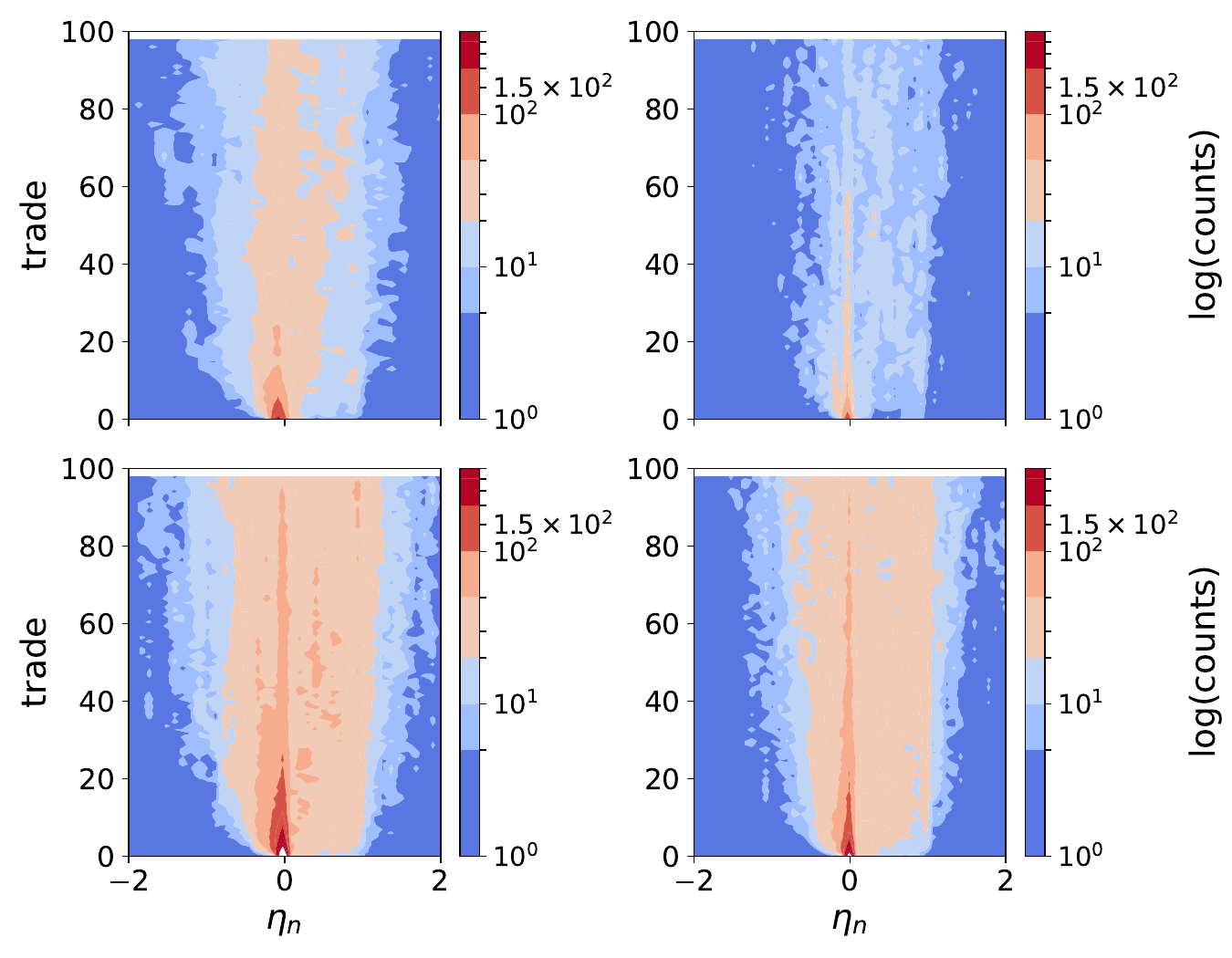}
	\end{subfigure}
	\caption{Contour plots of the recovery ratio (referred
		to as recovery rate in Ref.~\cite{braun2018impact}) dependent on the
		trade number for flash crashes (left) and flash spikes (right), in 2014 (top) and 2021 (bottom) on a logarithmic scale.}
	\label{Recover 2014_Log}
\end{figure}

\begin{figure}[htbp]
	\centering
	\begin{subfigure}[b]{1.0\textwidth}
		\centering
		\includegraphics[width=1.0\textwidth]{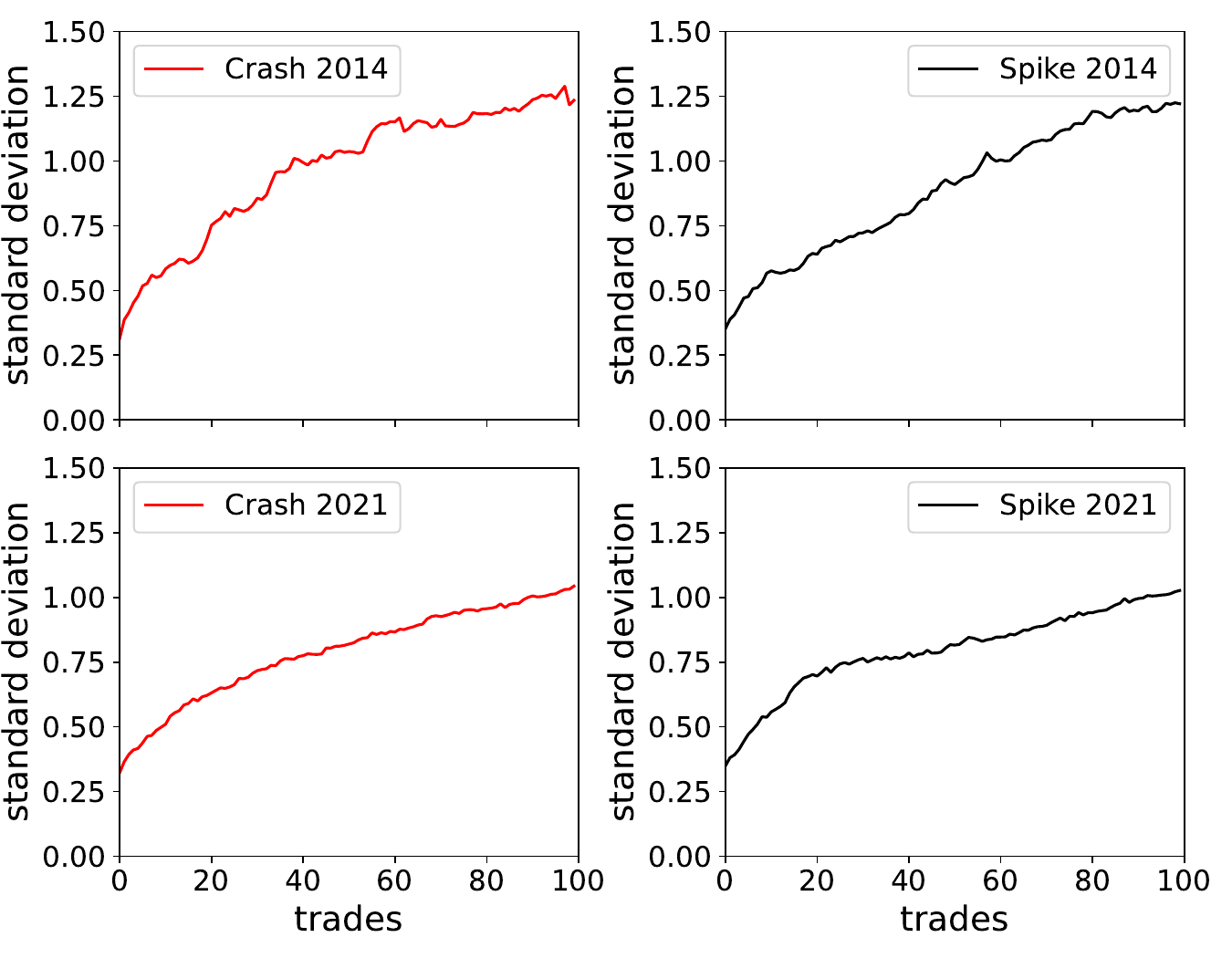}
	\end{subfigure}
	\caption{Standard deviations of the recovery ratio $\eta_n$ as a function of the number of trades, for flash crashes (left) and flash spikes (right) in 2014 (top) and 2021 (bottom).}
	\label{Prob 2014_SD}
\end{figure}

For each UEE in 2007/2008, 2014 and 2021 we show the recovery ratio
for the first 50 to 100 trades after the UEE
in Figs.~\ref{Recover 2014} and \ref{Recover 2014_Log} as
contour plots on a linear and logarithmic scale, respectively. 
The corresponding standard deviations of the recovery ratios,
computed across all stocks for a fixed number of trades,
are shown in Fig.~\ref{Prob 2014_SD}.
Clearly, for most of the flash crashes and flash spikes the prices at the end of a UEE stay on the extremum reached during the UEE.
Importantly, a few flash crashes and flash spikes recover immediately after one trade. Furthermore, there are only a few flash crashes and flash spikes that generate an ``aftershock'', \textit{i.\,e.} the price continues to rise for flash spikes and fall for flash crashes after the UEE leading to a negative value for the recovery ratios. 
Moreover, most of the recovery ratios take values in the interval $0\leq \eta_n \leq 1$. 


\begin{figure}[htbp]
	\centering
		\begin{subfigure}[b]{0.5\textwidth}
		\centering
		\includegraphics[width=1.0\textwidth]{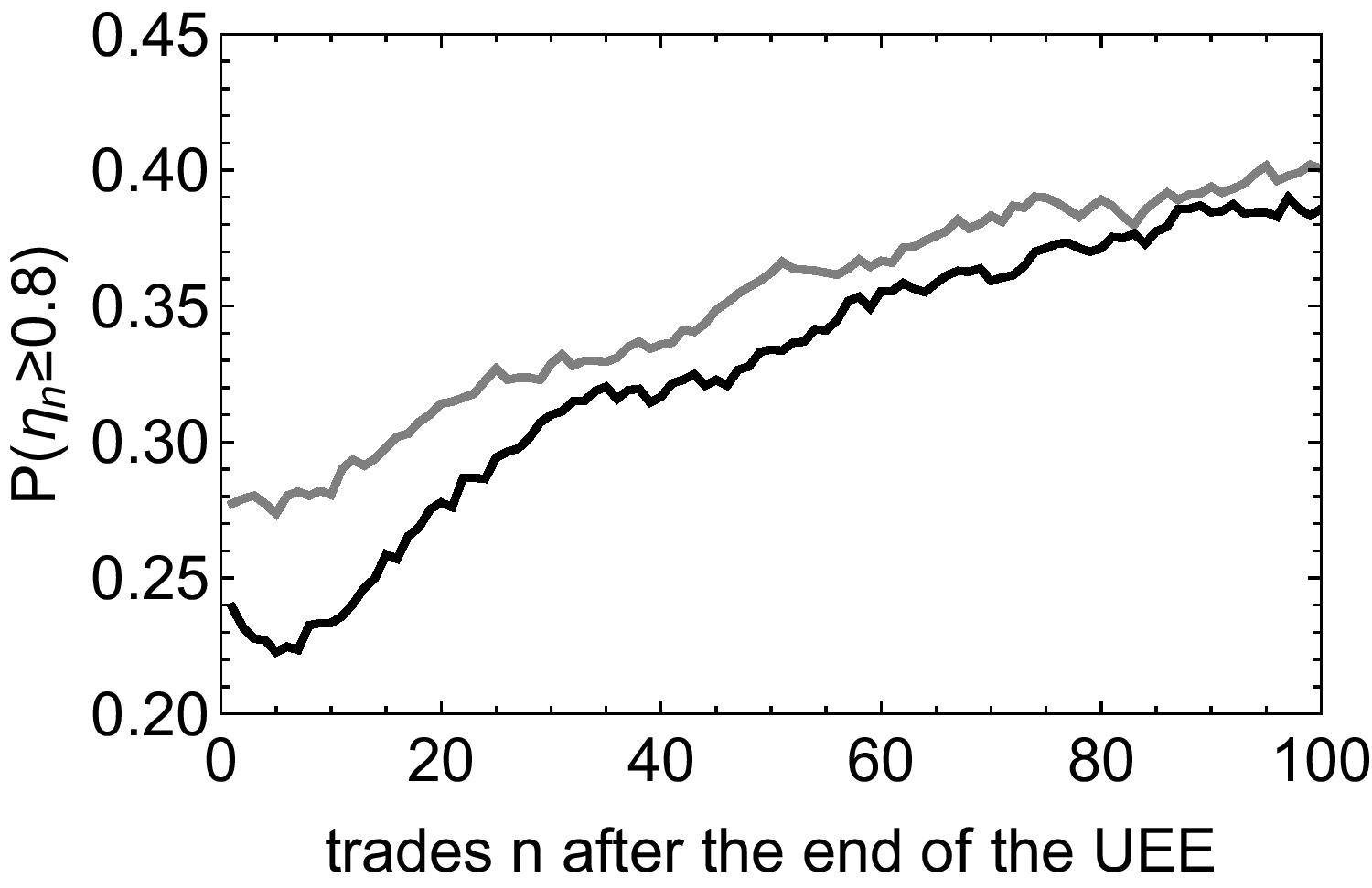}

	\end{subfigure}%
	\hfill
	\begin{subfigure}[b]{0.5\textwidth}
		\centering
		\includegraphics[width=1.0\textwidth]{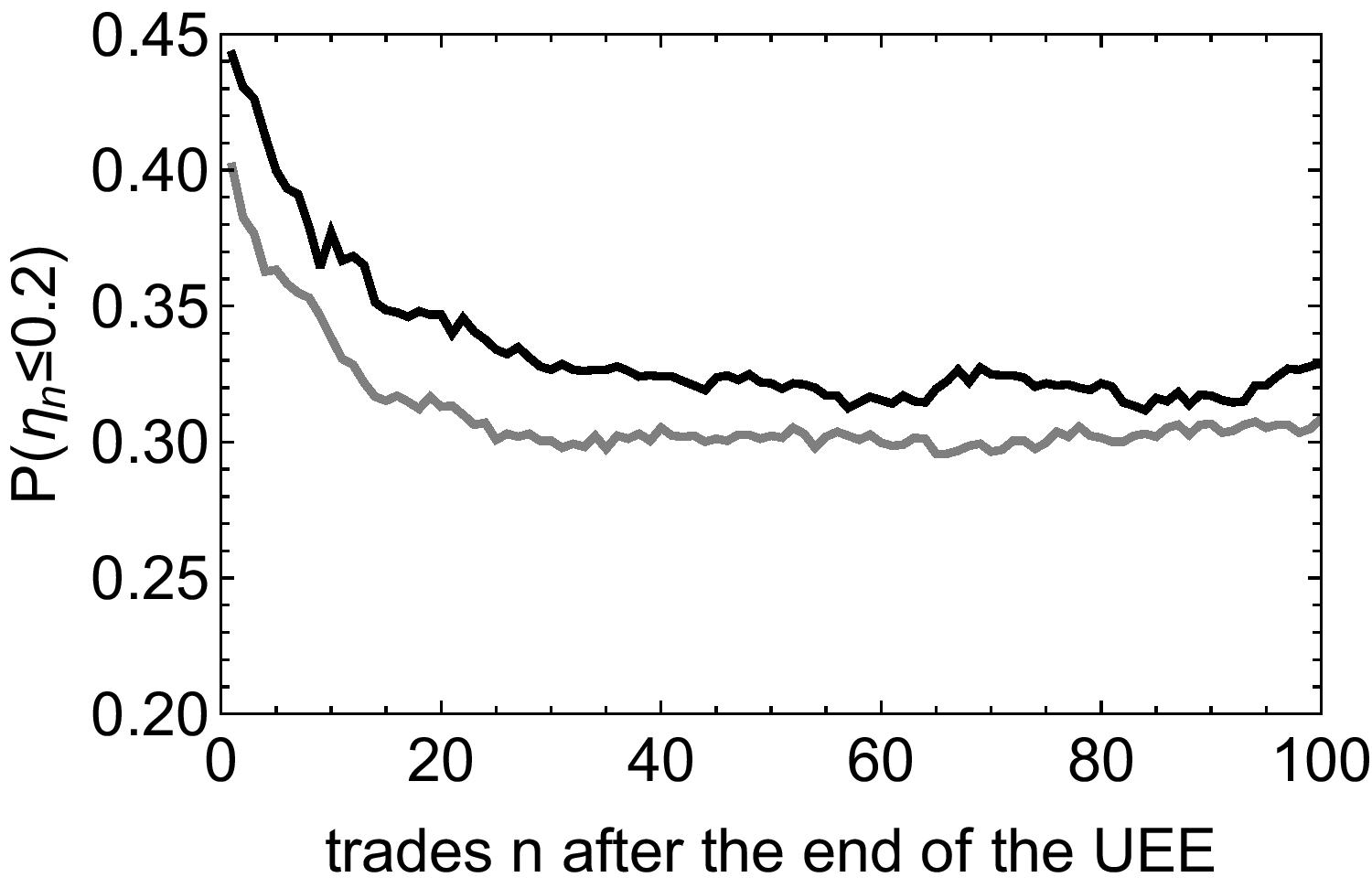}

	\end{subfigure}\\
	\begin{subfigure}[b]{1.0\textwidth}
		\centering
		\includegraphics[width=1.0\textwidth]{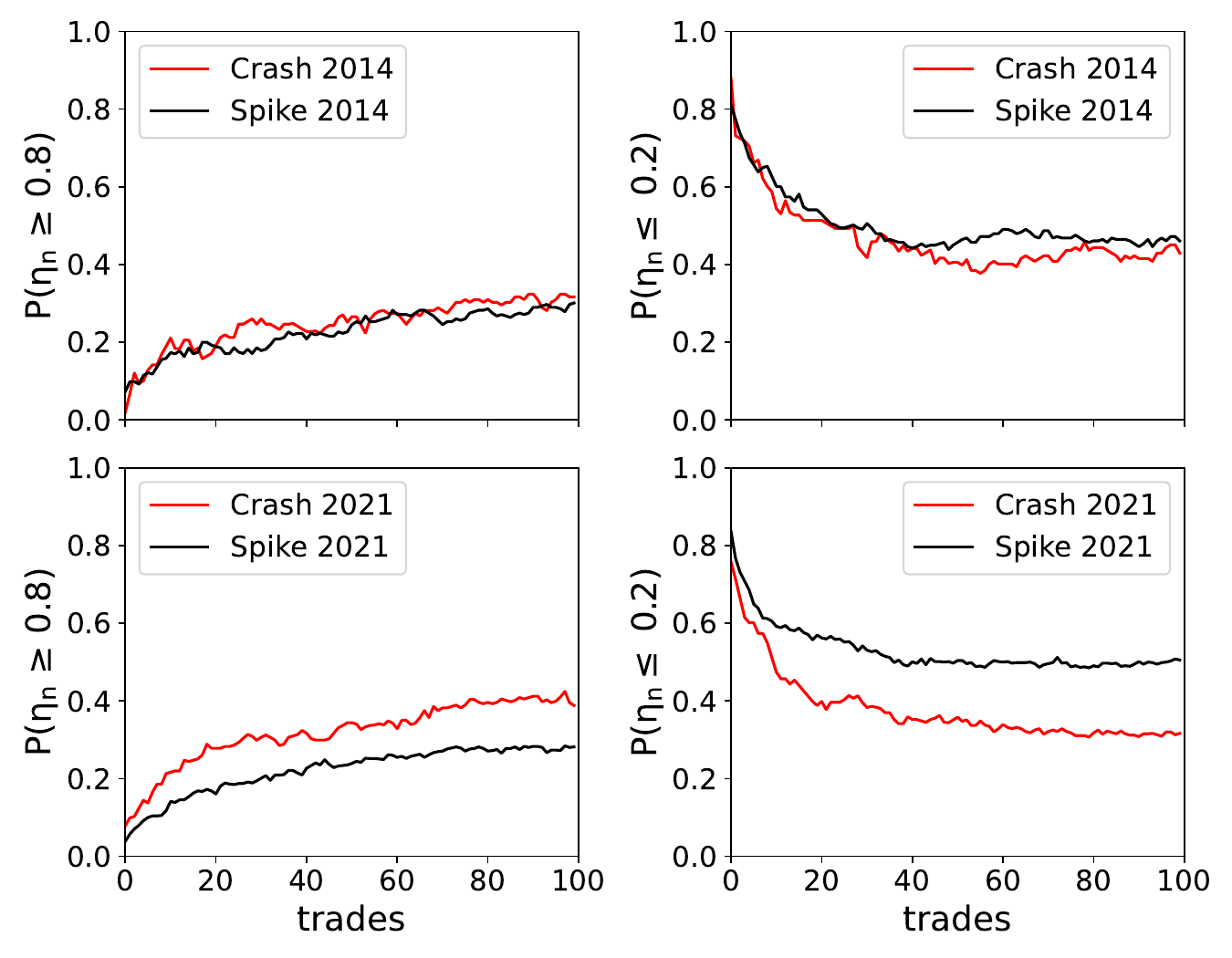 }
	\end{subfigure}
	\caption{Relative frequencies to have a recovery ratio of $\eta_n  \geq 0.8$ (left) and  $\eta_n  \leq 0.2$ (right) for 2007/2008 (top), 2014 (middle) and 2021 (bottom) on a linear scale. In the top figure, flash crashes are colored black, flash spikes gray. The top figure is taken from Ref.~\cite{braun2018impact}.}
	\label{Prob 2014}
\end{figure}

\begin{figure}[htbp]
	\centering
	\begin{subfigure}[b]{1.0\textwidth}
		\centering
		\includegraphics[width=1.0\textwidth]{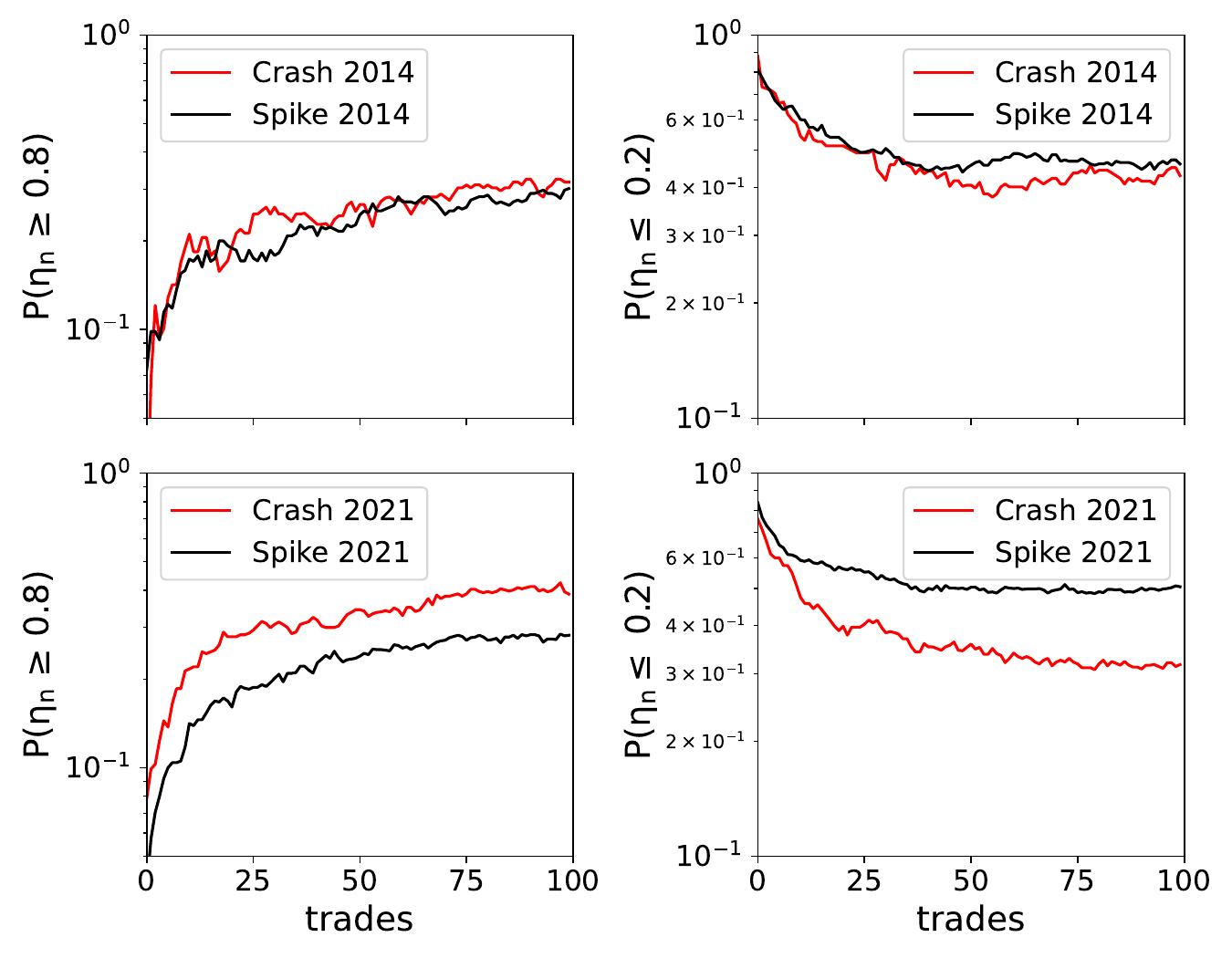}
	\end{subfigure}
	\caption{Relative frequencies to have a recovery ratio of $\eta_n  \geq 0.8$ (left) and  $\eta_n  \leq 0.2$ (right) for 2014 (top) and 2021 (bottom) on a logarithmic scale.}
	\label{Prob 2014_Log}
\end{figure}

For the sake of clarity, we work out relative frequencies of events for a fixed $n$ where $\eta_n$ is larger than (or equal to) $0.8$ and where $\eta_n$ is smaller than (or equal to) $0.2$.
In Figs.~\ref{Prob 2014} and \ref{Prob 2014_Log}, these relative frequencies are displayed for $1 \leq n \leq 100$ on a linear and logarithmic scale, respectively. 
For $P(\eta_n  \geq 0.8)$, \textit{i.\,e.} for the relative frequencies  for almost recovery,
flash crashes and flash spikes are similar in 2014. 
The relative frequencies  for almost recovery for flash crashes and flash spikes is faster in the years 2007/2008 for the very first trades.
For $P(\eta_n  \leq 0.2)$, \textit{i.e.} the relative frequencies  for almost no recovery, both flash crashes and flash spikes are very likely not to recover at the beginning. The relative frequencies  of flash crashes is significantly larger than for flash spikes in 2007/2008. 
In 2021, the flash spikes do not recover as well as the flash crashes.
The relative frequencies  for almost no recovery afterwards decreases and remains largely constant for $50\leq n \leq 100$ for all years.
Within the first 50 to 100 trades after a UEE it is determined whether a stock will recover almost completely or only incompletely from the shock caused by a UEE.

\section{\label{sec:Conlcusion}Conclusions}

We carried out a large-scale empirical analysis and observed stable empirical patterns in the formation and after-effects of UEEs, which hint at a certatin degree of universal behavior.
To uncover it, we analyzed various years independently.
As in a previous analysis of our group for 2007/2008~\cite{braun2018impact}, we found that one return in the quotes during a UEE often almost fulfills the UEE criterion. %
Our analysis extends the study of Ref.~\cite{braun2018impact}.
To make a comparison of the results for the years 2014 and 2021 with the empirical analysis of Ref.~\cite{braun2018impact} for 2007 feasible, we employed as a criterion of 2--seconds for the duration of UEEs. In addition, because the resolution of the 2014 and 2021 data is better, we also applied a 1.5--second criterion. Importantly, the number of UEEs is almost the same within less than 12\,\%. The statistical characteristics do not change much either. This indicates that the statistical properties of UEEs are fairly stable with respect to variations in the event duration criterion.

Most of the UEEs occur right after the pre-- and after--market trading. In these time periods, the liquidity builds up or decreases again~\cite{Krause_2022}.
Both of these time periods are dominated by cancelations of limit orders.
As the occurence of UEEs is a typical pattern for the years 2014 and 2021, it stands to reason that a large number of UEEs have similar underlying mechanisms.
The liquidity status of the order book due to cancelations of limit orders is probably a crucial factor here. 
Gaps in the order book, which can be caused by cancelations of limit orders, can lead to larger return jumps in the midpoints~\cite{Farmer_2004}.
From this point of view, it does not matter whether HFTs, other algorithmic traders or human traders are responsible for UEEs, as all types of traders can cancel limit orders and trigger larger price jumps through market orders.
Presumably, these types of traders after the pre-- and after--market trading play a minor role, as cancelations within this time periods seem to be a generic trading pattern.
It is reasonable to assume that, even during a trading day, the underlying mechanism for many UEEs may be due to brief liquidity shortages, initiated by cancelations of limit orders and triggered by market orders.
We collected evidence for this mechanism, but we do not exclude other mechanisms for UEEs.

In line with this interpretation, UEEs occur predominantly in periods of low trading activity and sparse liquidity, where the reduced order flow limits the extent to which rapid reactive trading could reinforce the initial movement. The short and isolated structure of the events is therefore consistent with a liquidity-driven rather than a reaction-driven mechanism. Further studies with access to full order-book information would be valuable to explore this mechanism in more detail.

It is interesting to compare our observations with the evolution of algorithmic trading~\cite{kissell2020algorithmic}.
We know that algorithmic trading has almost completely replaced human traders. For instance, 92\% of the stock volume  in the US stock markets was traded by algorithmic traders inclusively HFTs in 2019. Electronic trading accounted for 99,9\% of the total trading volume~\cite{kissell2020algorithmic}, \textit{i.e.} brokers (human traders) traded 8\% of the total volume via a computer trading system.
\begin{figure}[t]
	\centering 	
	\includegraphics[width=1.0\linewidth]{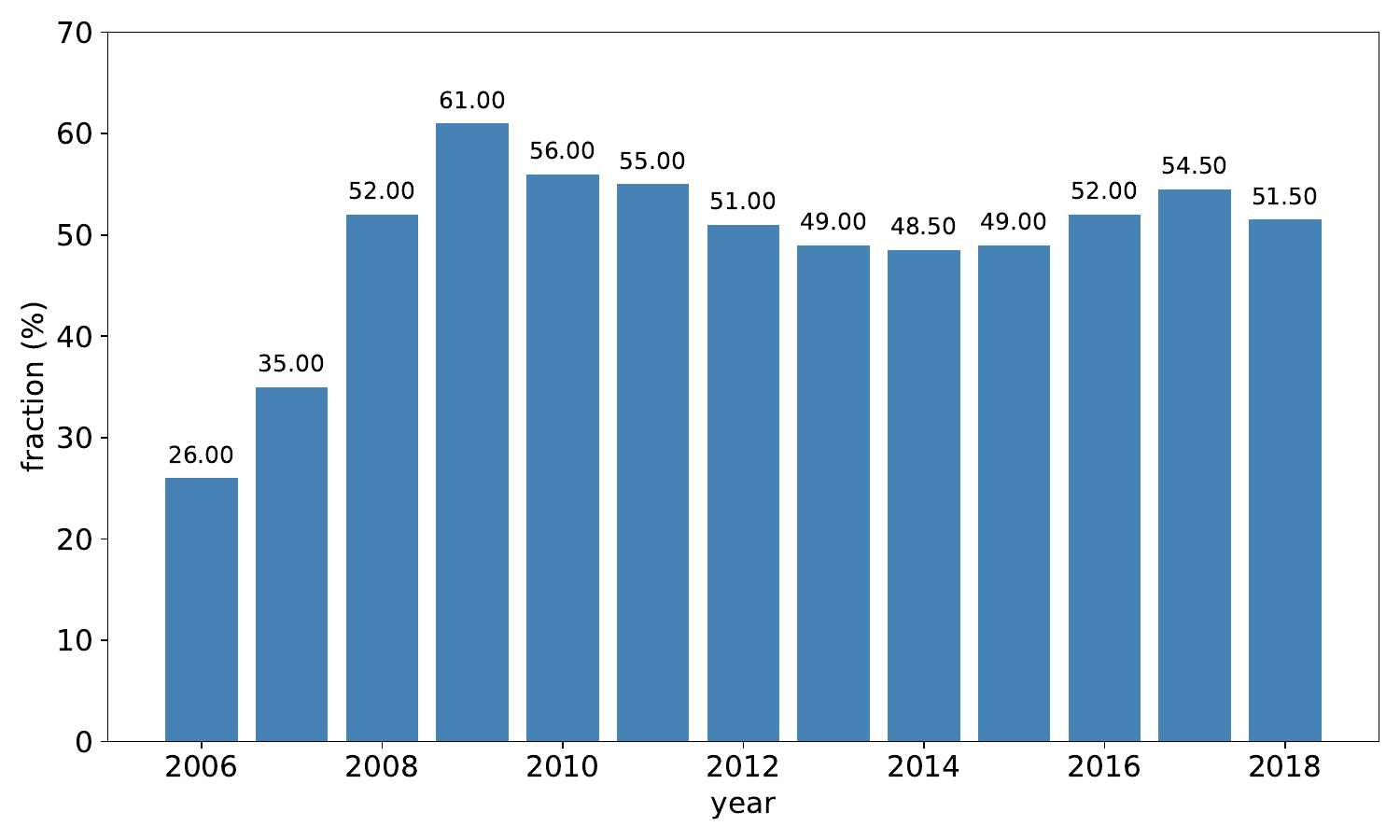}
	\caption{Percentage of HFT trading in total US stock volume.  Estimates from TABB Group, reported by Franklin Templeton. Taken from Ref.~\cite{Bauer_2022}.}
	\label{fig:TradersTotalVolume}
\end{figure}
For the US stock markets, HFTs account for %
26\% to 61\%~\cite{gerig2012highfrequencytradingsynchronizesprices,Ersan_2021,gkasiorkiewicz2021innovation,Zaharudin_2022,Bauer_2022} of algorithmic trading in terms of the total market volume in a time period from 2006 to 2018, see Fig.~\ref{fig:TradersTotalVolume}. These numbers may vary depending on the literature source, year and definition of HFTs~\cite{kissell2020algorithmic}. During the financial crisis of 2008/2009, the share of HFTs peaked. %
The interactions between human traders and HFTs or the interactions between HFTs and other algorithmic traders do not appear to be responsible for the emergence of UEEs.
Hence, our findings are independent of the composition of traders in the financial markets, \textit{i.e.} independent of this non--stationary aspect of trading.

The larger returns in the quotes appear to be not necessarily directly responsible for the formation of UEEs, as these only appear after the start of the UEE.
Rather, they might be important for the recovery process of a UEE.
For all four years, we observe stocks with an almost full recovery after a UEE. The almost full recovery will be due to the state of the order book with less fluctuating liquidity after a UEE hinting at a more stable state of the order book.
Moreover, we find that the qualitative shape of the recovery patterns is remarkably robust across years with very different levels of market-wide volatility (2007/2008, 2014, 2021), indicating that the dimensionless recovery ratio is largely insensitive to broad volatility regimes.
Even though there are similarities in the shape of the recovery ratio, the market sentiment causes non--stationary variations over the years.
For instance, one reason for the difference between flash crashes and flash spikes in 2021 could have been the general stock market situation in that year. In spite of the short squeeze
in 2021, the stock market year 2021 recovered from the COVID--19 pandemic \cite{WSJ,CNBC}. Traders may have tended to continue trading at higher price levels
because it was assumed that the prices of stocks would continue to rise.

\newpage

\section*{Acknowledgement}

We thank Benjamin Köhler and Cedric Schuhmann for fruitful discussions.

\clearpage

\bibliography{Lit.bib}

\clearpage

\newpage

\begin{appendices}
\markboth{Appendix}{Appendix}

\section{\label{Appendix}Appendix}

\begin{table*}
	\linespread{0.5}
	\caption{100 stocks from ten economic sectors for 2014.} 
	\begin{center}
		\begin{footnotesize}
			\begin{tabular}{llcll} 
				\toprule
				\multicolumn{2}{l}{\textbf{Industrials} (I)} &~~& \multicolumn{2}{l}{\textbf{Financials} (F)}  \vspace{0.1cm} \\
				Symbol	&Company				 	&	&Symbol		&Company						\\
				\midrule
				FLR		&Fluor Corp. (New)				&	&CME		&CME Group Inc.					\\
				LMT		&Lockheed Martin Corp.			&	&GS			&Goldman Sachs Group				\\
				FLS		&Flowserve Corporation			&	&ICE			&Intercontinental Exchange Inc.		\\
				PCP		&Precision Castparts			&	&AVB		&AvalonBay Communities				\\
				LLL		&L-3 Communications Holdings	&	&BEN		&Franklin Resources					\\
				UNP		&Union Pacific					&	&BXP		&Boston Properties				 	\\
				{AAL}		&{American Airlines Group Inc.} &	&SPG		&Simon Property 	Group  Inc	 		\\
				FDX		&FedEx Corporation				&	&VNO		&Vornado Realty Trust				\\
				GWW	&Grainger (W.W.) Inc.			&	&PSA		&Public Storage					\\
				GD		&General Dynamics				&	&MTB		&M$\&$T Bank Corp.				\\
						&				&	&JPM		&JP Morgan Chase				\\
				\midrule
				\multicolumn{2}{l}{\textbf{Health Care} (HC)} &~& \multicolumn{2}{l}{\textbf{Materials} (M)} 	\vspace{0.1cm}  \\
				Symbol	&Company					&	&Symbol		&Company						\\
				\midrule
				ISRG	&Intuitive Surgical Inc.			&	 &X			&United States Steel Corp.			\\
				BCR		&Bard (C.R.) Inc.				&	 &MON		&Monsanto Co.						\\
				BDX		&Becton  Dickinson				&	 &CF			&CF Industries Holdings Inc			\\
				{BIIB}	&{Biogen Idec Inc.}.				&	 &FCX		&Freeport-McMoran Cp $\&$ Gld 		\\
				JNJ		&Johnson $\&$ Johnson			&	&APD		&Air Products $\&$ Chemicals			\\
				LH		&Laboratory Corp. of America Holding & &PX		&Praxair  Inc.						\\
				ESRX	&Express Scripts				&	&VMC		&Vulcan Materials					\\
				CELG	&Celgene Corp.				&	&{DOW}		&{DOW Chemicals Co. Com.} \\
				ZMH		&Zimmer Holdings				&	&NUE		&Nucor Corp.						\\
				AMGN	&Amgen						&	&PPG		&PPG Industries					\\
				\midrule
				\multicolumn{2}{l}{\textbf{Consumer	Discretionary} (CD)} &~& \multicolumn{2}{l}{\textbf{Energy} (E)}  \vspace{0.1cm}  \\
				Symbol	&Company					&	&Symbol		&Company					\\
				\midrule
				{GHC}	&{Graham Holdings}			&	&RIG		&Transocean Inc. (New)				\\
				AZO		&AutoZone Inc.					&	&APA		&Apache Corp.					\\
				SHLD	&Sears Holdings Corporation		&	&EOG		&EOG Resources					\\
				WYNN	&Wynn Resorts Ltd.				&	&DVN		&Devon Energy Corp.				\\
				AMZN	&Amazon Corp.				&	&HES		&Hess Corporation					\\
				WHR	&Whirlpool Corp.			&	&XOM		&Exxon Mobil Corp.					\\
				VFC		&V.F. Corp.					&	&SLB		&Schlumberger Ltd.					\\
				APOL	&Apollo Group					&	&CVX		&Chevron Corp.					\\
				NKE		&NIKE Inc.					&	&COP		&ConocoPhillips					\\
				MCD		&McDonald's Corp.				&	&OXY		&Occidental Petroleum				\\
				\midrule
				\multicolumn{2}{l}{\textbf{Information Technology} (IT)} &~& \multicolumn{2}{l}{\textbf{Consumer Staples} (CS)}  \vspace{0.1cm} \\
				Symbol	&Company					&	&Symbol		&Company						\\
				\midrule
				GOOG	&Google Inc.					&	&BUD		&Anheuser-Busch					\\
				MA		&Mastercard Inc.			&	&PG			&Procter $\&$ Gamble				\\
				AAPL	&Apple Inc.					&	&CL			&Colgate-Palmolive					\\
				IBM		&International Bus. Machines		&	&COST		&Costco Co.						\\
				MSFT	&Microsoft Corp.			& 	&WMT		&Wal-Mart Stores					\\
				CSCO	&Cisco Systems				&	&PEP		&PepsiCo Inc.						\\
				INTC		&Intel Corp.					&	&LO			&Lorillard Inc.						\\
				QCOM	&QUALCOMM Inc.				&	&{MO}		&{Altria Group Inc.}						\\
				CRM		&Salesforce Com Inc. 			&	&GIS		&General Mills						\\
				{SUNE}		&{Sunedison Inc. Com.}		&	&KMB		&Kimberly-Clark					\\
				\midrule
				\multicolumn{2}{l}{\textbf{Utilities} (U)} &~& \multicolumn{2}{l}{\textbf{Telecommunications Services} (TS)} \vspace{0.1cm}  \\
				Symbol	&Company					&	&Symbol		&Company						\\
				\midrule
				ETR		&Entergy Corp.					&	&T			&AT$\&$T Inc.						\\
				EXC		&Exelon Corp.					&	&VZ			&Verizon Communications			\\
				{DUK}		&{Duke Energy Corp. Com.} 		&	&{FB}			&{Facebook Inc. Com.}				\\
				FE		&FirstEnergy Corp.				&	&AMT		&American Tower Corp.				\\
				{NEE}		&{NextEra Energy Inc. Com.}					&	&CTL 		&Century Telephone					\\
				SRE		&Sempra Energy				&	&S			&Sprint Nextel Corp.					\\
				STR		&Questar Corp.					&	&Q			&Qwest Communications  Int 			\\
				TEG		&Integrys Energy Group Inc. 		&	&WIN		&Windstream Corporation			\\
				EIX		&Edison Int'l					&	&FTR		&Frontier Communications			\\
				{SO}		&{Southern Com.}				&	&			&									\\
				\bottomrule
			\end{tabular}
		\end{footnotesize}
	\end{center}
	\label{tab:stocks_2014}
\end{table*}

\begin{table*}
	\linespread{0.5}
	\caption{100 stocks from ten economic sectors for 2021.} 
	\begin{center}
		\begin{footnotesize}
			\begin{tabular}{llcll} 
				\toprule
				\multicolumn{2}{l}{\textbf{Industrials} (I)} &~~& \multicolumn{2}{l}{\textbf{Financials} (F)}  \vspace{0.1cm} \\
				Symbol	&Company				 	&	&Symbol		&Company						\\
				\midrule
				FLR		&Fluor Corp. (New)				&	&CME		&CME Group Inc.					\\
				LMT		&Lockheed Martin Corp.			&	&GS			&Goldman Sachs Group				\\
				FLS		&Flowserve Corporation			&	&ICE			&Intercontinental Exchange Inc.		\\
				{BA}		&{The Boeing Company}			&	&AVB		&AvalonBay Communities				\\
				{LHX}		&{L3 Harris Technologies}	&	&BEN		&Franklin Resources					\\
				UNP		&Union Pacific					&	&BXP		&Boston Properties				 	\\
				{AAL}		&{American Airlines Group Inc.} &	&SPG		&Simon Property 	Group  Inc	 		\\
				FDX		&FedEx Corporation				&	&VNO		&Vornado Realty Trust				\\
				GWW	&Grainger (W.W.) Inc.			&	&PSA		&Public Storage					\\
				GD		&General Dynamics				&	&MTB		&M$\&$T Bank Corp.				\\
				&				&	&JPM		&JP Morgan Chase				\\
				\midrule
				\multicolumn{2}{l}{\textbf{Health Care} (HC)} &~& \multicolumn{2}{l}{\textbf{Materials} (M)} 	\vspace{0.1cm}  \\
				Symbol	&Company					&	&Symbol		&Company						\\
				\midrule
				ISRG	&Intuitive Surgical Inc.			&	 &X			&United States Steel Corp.			\\
				{MRNA}		&{Moderna Inc.}				&	 &{VALE}		&{Vale S.A.}						\\
				BDX		&Becton  Dickinson				&	 &CF			&CF Industries Holdings Inc			\\
				{BIIB}	&{Biogen Idec Inc.}.				&	 &FCX		&Freeport-McMoran Cp $\&$ Gld 		\\
				JNJ		&Johnson $\&$ Johnson			&	&APD		&Air Products $\&$ Chemicals			\\
				LH		&Laboratory Corp. of America Holding & &{LIN}		&{Linde plc.}						\\
				{CI}	&{Cigna Corp}				&	&VMC		&Vulcan Materials					\\
				{BMY}	&{Bristol-Meyers Squibb}				&	&{DOW}		&{DOW Chemicals Co. Com.} \\
				{ZBH}		&{Zimmer Biomet Holdings Inc.}				&	&NUE		&Nucor Corp.						\\
				AMGN	&Amgen						&	&PPG		&PPG Industries					\\
				\midrule
				\multicolumn{2}{l}{\textbf{Consumer	Discretionary} (CD)} &~& \multicolumn{2}{l}{\textbf{Energy} (E)}  \vspace{0.1cm}  \\
				Symbol	&Company					&	&Symbol		&Company					\\
				\midrule
				{GHC}	&{Graham Holdings}			&	&RIG		&Transocean Inc. (New)				\\
				AZO		&AutoZone Inc.					&	&APA		&Apache Corp.					\\
				{TSLA}	&{Tesla Inc}		&	&EOG		&EOG Resources					\\
				WYNN	&Wynn Resorts Ltd.				&	&DVN		&Devon Energy Corp.				\\
				AMZN	&Amazon Corp.				&	&HES		&Hess Corporation					\\
				WHR	&Whirlpool Corp.			&	&XOM		&Exxon Mobil Corp.					\\
				VFC		&V.F. Corp.					&	&SLB		&Schlumberger Ltd.					\\
				{BABA}	&{Alibaba}					&	&CVX		&Chevron Corp.					\\
				NKE		&NIKE Inc.					&	&COP		&ConocoPhillips					\\
				MCD		&McDonald's Corp.				&	&OXY		&Occidental Petroleum				\\
				\midrule
				\multicolumn{2}{l}{\textbf{Information Technology} (IT)} &~& \multicolumn{2}{l}{\textbf{Consumer Staples} (CS)}  \vspace{0.1cm} \\
				Symbol	&Company					&	&Symbol		&Company						\\
				\midrule
				GOOG	&Google Inc.					&	&BUD		&Anheuser-Busch					\\
				MA		&Mastercard Inc.			&	&PG			&Procter $\&$ Gamble				\\
				AAPL	&Apple Inc.					&	&CL			&Colgate-Palmolive					\\
				IBM		&International Bus. Machines		&	&COST		&Costco Co.						\\
				MSFT	&Microsoft Corp.			& 	&WMT		&Wal-Mart Stores					\\
				CSCO	&Cisco Systems				&	&PEP		&PepsiCo Inc.						\\
				INTC		&Intel Corp.					&	&{BTI}	&{Britisch American Tobacco}						\\
				QCOM	&QUALCOMM Inc.				&	&{MO}		&{Altria Group Inc.}						\\
				CRM		&Salesforce Com Inc. 			&	&GIS		&General Mills						\\
				{NVDA}		&{Nvidia Corp.}		&	&KMB		&Kimberly-Clark					\\
				\midrule
				\multicolumn{2}{l}{\textbf{Utilities} (U)} &~& \multicolumn{2}{l}{\textbf{Telecommunications Services} (TS)} \vspace{0.1cm}  \\
				Symbol	&Company					&	&Symbol		&Company						\\
				\midrule
				ETR		&Entergy Corp.					&	&T			&AT$\&$T Inc.						\\
				EXC		&Exelon Corp.					&	&VZ			&Verizon Communications			\\
				{DUK}		&{Duke Energy Corp. Com.} 		&	&{FB}			&{Facebook Inc. Com.}				\\
				FE		&FirstEnergy Corp.				&	&AMT		&American Tower Corp.				\\
				{NEE}		&{NextEra Energy Inc. Com.}					&	&{LUMN} 		&{Lumen Technologies}				\\
				SRE		&Sempra Energy				&	&{TMUS}			&{T-Mobile US Inc.}					\\
				{D}		&{Dominion Energy Inc.}					&	&{IQV}			&{IQVIA Holdings Inc.} 			\\
				{WEC}		&{WEC Energy Group} 		&	&{NFLX}		&{Netflix Inc.}			\\
				EIX		&Edison Int'l					&	&{BIDU}		&{Baidu Inc.}			\\
				{SO}		&{Southern Com.}				&	&			&									\\
				\bottomrule
			\end{tabular}
		\end{footnotesize}
	\end{center}
	\label{tab:stocks_2021}
\end{table*}

\end{appendices}

\end{document}